\documentclass[twoside,twocolumn,pra,aps,showpacs]{revtex4}
\usepackage{amsmath,amssymb,epsfig}
\begin{document}
%\preprint{\today}
%\draft
%
%%%%%%%%%%%%%%%%%%%%%%%%%%%%%%%%% TITLE PAGE
%
\title{Entangling ions in arrays of microscopic traps}
\author{T. Calarco$^{1,2}$, J.I. Cirac$^1$, and P. Zoller$^1$}
\affiliation{$^1$Institut f\"ur Theoretische Physik, Universit\"at Innsbruck,
Technikerstra\ss e 25, A--6020 Innsbruck, Austria\\
$^2$ECT*, European Centre for Theoretical Studies in Nuclear Physics
and Related Areas,
Villa Tambosi, Strada delle Tabarelle 286, I--38050 Villazzano (Trento), Italy
}
\date{\today}
%
%%%%%%%%%%%%%%%%%%%%%%%%%%%%%%%%% ABSTRACT
%
\begin{abstract}
We consider a system of particles in an array of microscopic
traps, coupled to each other via electrostatic interaction, and
pushed by an external state-dependent force. We show how to
implement a two-qubit quantum gate between two such particles
with a high fidelity.
\end{abstract}
%
%%%%%%%%%%%%%%%%%%%%%%%%%%%%%%%%% PACS NUMBERS
%
\maketitle
%%%%%%%%%%%%%%%%%%%%%%%%%%%%%%%%% PAPER CONTENT
%
%\narrowtext
%
\section{Introduction}

The possibilities offered by quantum mechanical systems for
efficient information processing have stimulated in recent years
the rise of an entirely new field of research \cite{Nielsen}.
Quantum protocols for secure communication over long distances
have been devised and demonstrated. Quantum algorithms for
efficient solution of problems believed to be intractable on
classical computers have been developed. However, while quantum
communication is already approaching the stage of real-world
applications, quantum computation remains still at a less
advanced level, as far as physical implementation is concerned.
Different systems are being proposed as candidates for this
purpose \cite{FdP}, but nobody can yet tell what will turn out to
be a viable solution. Indeed, in a few cases quantum computation
building blocks --~single- and two-qubit operations~-- have been
already demonstrated experimentally. In principle, these
ingredients are universal -- they are sufficient to build an
arbitrary unitary transformation over $N$ qubits (i.e., any
quantum computation). But in order to perform useful computations
in a real environment inducing decoherence, fault tolerance is
also required. This implies e.g. nested redundant coding for
real-time error correction \cite{fault}, and requires an error
probability for elementary operations below a certain threshold
(of the order of $10^{-4}$). Hence the need for new proposals,
allowing for handling a bigger number of qubits at a lower
decoherence rate and with faster and more reliable gate
operations -- in a word, enabling scalability of the system.

We propose to use quantum optical systems in periodic microscopic
potentials. This is meant to combine the good isolation and
precise control by laser fields, achievable in quantum optics,
with the ability --~usually associated with semiconductor
technology~-- of manufacturing periodic structures to generate
modulated fields on a microscopic scale. The general concept of
our proposal is to encode the logical states of each qubit into
two internal states of a particle (neutral atom or ion).
Single-qubit operations are obtained as Rabi rotations by
applying resonant laser fields. Two-qubit gates are performed by
inducing a state-dependent interaction over a certain time,
making the particles acquire a conditional phase shift depending
on their logical states. These, however, in a real situation are
coupled to other external degrees of freedom. This can lead to
different kinds of imperfections. On one hand, the external state
after gate operation may not be exactly the same as before. On the
other hand, the conditional phase shift will also depend on the
external state: if this is mixed, only an imprecise phase
determination will be possible. These facts affect the gate
fidelity, which is defined by comparing the desired effect of the
gate with the actual evolution that can be obtained in the
laboratory. We already proposed several schemes, based on
different interactions -- collisional interactions between
neutral atoms in optical lattices \cite{OptLatt} and magnetic
microtraps \cite{MagnTrap}, or dipole-dipole interactions between
Rydberg-excited atoms \cite{Rydb}. Here we deal with
electrostatic interaction between ions \cite{ion95,ion2000} in
arrays of microscopic traps \cite{DeVoe}.

In this paper we describe the conditional dynamics for two charged
particles, trapped in separate harmonic wells, interacting via
electrostatic repulsion and under the influence of an external
state-dependent force, which can be generated e.g. by an
off-resonant laser standing wave \cite{ion2000}. The goal is to
implement a phase gate between the two qubits
(Sec.~\ref{section:defgate}), i.e. to transform their initial
state by inducing a certain phase onto each of its components. The
ideal transformed state so defined has to be compared to the one
that can be obtained by means of a realistic Hamiltonian,
coupling the particles' internal and external degrees of freedom
(Sec.~\ref{section:cond}). To this aim, we consider first a
one-dimensional classical model for the motion
(Sec.~\ref{section:class}). We solve the equations of motion for
each combination of logical states separately, and define in each
case a two-particle phase as the integral over time of the
interaction energy. These phases define the evolved internal
state, to be eventually compared with the ideal state we aim to
obtain. The fidelity of gate operation is evaluated, in this
classical model, as the overlap between the real and the ideal
state, averaged over different starting conditions according to a
thermal probability distribution for the initial oscillation
energies. The outcome is a series expansion for the temperature
dependence of the fidelity, of which we give explicitly the first
terms. The full three-dimensional quantum-mechanical calculation
follows the same path (Sec.~\ref{section:quantum}), except for a
couple of points. First, the two-particle phases are now given
directly by the Schr\"odinger equation and not defined ad hoc as
before. Second, the fidelity is evaluated by tracing out the
external variables of a density matrix representing a mixed
thermal quantum state. We calculate perturbative corrections
arising from multipole terms in the Coulomb potential, and show
how to suppress lowest-order corrections to the fidelity by means
of an intermediate $\pi$ rotation on the qubits, thus achieving an
improvement by several orders of magnitude.

\section{A quantum phase gate}
\label{section:defgate}

We want to implement quantum logic between particles stored in an
array of microscopic traps. The qubits' logical states
$|0\rangle$ and $|1\rangle$ are encoded into particles' internal
states. One basic building block towards multi-qubit entanglement
operations is the phase gate between two qubits -- a
transformation which rotates by a certain phase just one
component of logical states:
\begin{equation}
\begin{array}{rcrl}
|0\rangle|0\rangle &\longrightarrow &|0\rangle|0\rangle&\\
|0\rangle|1\rangle &\longrightarrow &|0\rangle|1\rangle&\\
|1\rangle|0\rangle &\longrightarrow &|1\rangle|0\rangle&\\
|1\rangle|1\rangle &\longrightarrow &e^{i\vartheta}|1\rangle|1\rangle &\!\!\!.
\end{array}
\label{defGate}
\end{equation}
When $\vartheta=\pi$, this is equivalent --~up to single qubit
rotations~-- to a Controlled-NOT gate. Ideally, this would be
accomplished by means of a state-dependent interaction of the form
\begin{equation}
H_{\rm int}=\Delta E(t)|1\rangle_1\langle 1|
\otimes|1\rangle_2\langle 1|,
\end{equation}
acting over a time $\tau$ such that
\begin{equation}
\int_0^\tau\Delta E(t')dt'=\vartheta.
\end{equation}
However, it is not straightforward to realize in practice an
interaction between two particles which couples only their
internal states -- other degrees of freedom, for instance the
motional ones, are likely to be affected. Therefore our goal is
to approximate the ideal transformation Eq.~(\ref{defGate}) by
means of a conditional dynamics for two particles, making them
acquire the phase $\vartheta$ if and only if they are both in the
internal state $|1\rangle$, and leaving eventually the external
degrees of freedom practically unaffected. This is described a
Hamiltonian of the form
\begin{equation}
\label{defHab}
H(t,\mathbf{x}_1,\mathbf{x}_2)=\sum_{\alpha,\beta=0}^1
H^{\alpha\beta}(t,\mathbf{x}_1,\mathbf{x}_2)\;
|\alpha\rangle_1\langle\alpha|\otimes|\beta\rangle_2\langle\beta|,
\end{equation}
where $\mathbf{x}_j$ denotes the external degrees of freedom of particle $j$, and
the explicit time dependence indicates that we can switch on and off a
suitable interaction in order to obtain the desired effect.
To evaluate the performance of our scheme,
we have to compare the case of an ideal gate, as given by Eq.~(\ref{defGate}),
with the gate that can be actually realized by the physical process described
by Eq.~(\ref{defHab}).
The figure of merit is the minimum fidelity $F$, given by
\begin{equation}
F=\min_{\chi}\;{\rm tr}_{\rm ext}\langle\chi'|\sigma'|\chi'\rangle,\label{defFid}
\end{equation}
where ${\rm tr}_{\rm ext}$ denotes the trace over the external
degrees of freedom,
$|\chi\rangle\equiv\sum_{\alpha\beta}c_{\alpha\beta}
|\alpha\rangle_1|\beta\rangle_2$ is a generic two-ion internal
state, $|\chi'\rangle$ is the state obtained from $|\chi\rangle$
via the transformation Eq.~(\ref{defGate}), and $\sigma'$ is the
total density matrix, including external degrees of freedom,
after the evolution dictated by the Hamiltonian
Eq.~(\ref{defHab}), starting from an initial state
\begin{equation}
\label{defsigma}
\sigma\equiv\rho_1(t_0)\otimes\rho_2(t_0)\otimes|\chi\rangle\langle\chi|,
\end{equation}
where $\rho_j(t_0)$ is the external state of particle $j$ at the initial time
$t_0$. Ideally, to achieve the optimal fidelity $F=1$, we
need that the external degrees of freedom factorize after the gate operation, and
that the evolution operator
\begin{equation}
\label{defUt}
U(t,t_0)\equiv T\exp\left\{-\frac i\hbar
\int_{t_0}^tH(t',\mathbf{x}_1,\mathbf{x}_2)dt'\right\}
\end{equation}
has the only effect to induce a two-particle phase
$\varphi^{\alpha\beta}$, depending on the internal state of both
ions, plus single-particle phases, due to the kinetic energy
associated with the trap displacement. The latter can be undone by
means of single-qubit rotations (see App.~\ref{appendix:single}),
leaving us with the gate phase
\begin{equation}
\label{defvartheta}
\vartheta=\varphi^{00}-\varphi^{01}-\varphi^{10}+\varphi^{11}.
\end{equation}
In a real situation, the starting point will be rather a mixed
state corresponding to a thermal distribution over the external
energy eigenstates. In other words, at nonzero temperatures there
will be a finite probability that each particle starts in an
excited motional state, leading in general to different phases,
which cannot be experimentally controlled and easily undone by
single-qubit rotations. Therefore the fidelity is expected to
decrease with temperature, as we are going to show quantitatively
in the next Sections, both in a classical model for the particles'
motion and in a fully quantum framework.

\section{Conditional dynamics}
\label{section:cond}

We consider $N$ ions, trapped at positions denoted by (c-numbers)
${\mathbf {{\bar
r}}}_i$ ($1\leq i\leq N$). For simplicity, we take the trapping
potentials for all ions to be harmonic, with the same frequency
$\omega$ along every spatial direction. Our results can be
straightforwardly generalized to inhomogeneous trap arrays with
anisotropic confinement. Moreover, each ion is assumed to be
subject to a time-varying force ${\mathbf F}_i(t)$, depending on
its internal state $\alpha_i\in\{0,1\}$ as in Eq.~(\ref{defHab}).
The Hamiltonian is
\begin{equation}
H=\sum_{i=1}^{N}H_i+\sum_{i< j}^{N}H_{ij},
\label{HamN}
\end{equation}
where
\begin{subequations}
\begin{eqnarray}
H_i&\equiv &\frac{{\mathbf p}_i^2}{2m}+\frac 12m\omega^2\left(
{\mathbf r}_i-{\mathbf {\bar r}}_i\right) ^2
-{\mathbf F}_i(t)\cdot {\mathbf r}_i,\label{defHi}\\
H_{ij}&\equiv &\frac{q_e^2}{4\pi \varepsilon _0}
\frac1{\left| {\mathbf r}_i-{\mathbf r}_j\right| }.\label{defHij}
\end{eqnarray}
\end{subequations}
In the following, we will focus on two-particle dynamics.
We assume the external force to have the same strength on both ions,
i.e. to depend only on the internal state of each particle:
\begin{equation}
{\mathbf F}_i(t)=\sum_{\alpha=0}^1|\alpha\rangle_i\langle\alpha|
\otimes\mathbf{F}^{\alpha}(t).
\end{equation}
We can rewrite the Hamiltonian in Eq.~(\ref{defHab}) as
\begin{eqnarray}
H^{\alpha_1\alpha_2}(t,\mathbf{x}_1,\mathbf{x}_2)&=&
\sum_{i=1}^{2}\left\{\frac{\mathbf{p}_{i}^{2}}{2m}+\frac{m\omega^{2}}{2}
\left[\mathbf{x}_i-\mathbf{x}^{\alpha_{i}}(t)\right]^2\right.\quad
\label{defHa1a2}\\
&&\mbox{}\left.-\frac{\mathbf{F}^{\alpha_i}(t)^2}{2}\right\}
+\frac{q_e^2}{4\pi \varepsilon _0}
\frac1{\left|\mathbf{d}+\mathbf{x}_2-\mathbf{x}_1\right|},\nonumber
\end{eqnarray}
where we have defined $\mathbf{x}_1\equiv{\mathbf r}_1+{\mathbf
d}/2$, $\mathbf{x}_2\equiv{\mathbf r}_2-{\mathbf d}/2$,
$\mathbf{x}^{\alpha}(t)\equiv\mathbf{F}^{\alpha}(t)/(m\omega^2)$,
and $\mathbf d$ is the equilibrium interparticle separation. We
assume the external force to act along its direction, and we
choose the simple state dependence $\mathbf
F^\alpha(t)\equiv\big(\alpha\hbar\omega{\cal
F}(t)/a_\omega,0,0\big)$. Here $a_\omega\equiv\sqrt{\hbar
/m\omega}$ is the quantum harmonic oscillator ground-state width.
Hence the adimensional quantity ${\cal F}(t)$ represents the
displacement, in units of $a_\omega$, induced by the force on the
trap minimum for ion $i$, if it is in internal state
$|1\rangle_i$: indeed, we have $\mathbf x^\alpha(t)=\alpha
a_\omega\big({\cal F}(t),0,0\big)$. With the above choice for the
state dependence of the force, the last term in the first row at
right-hand side of Eq.~(\ref{defHa1a2}) will not contribute to
the gate phase Eq.~(\ref{defvartheta}), since the corresponding
terms cancel each others in the sum
$\sum_{\alpha,\beta}(-1)^{\alpha+\beta}\varphi^{\alpha\beta}$.
Since the interaction only depends on the distance between the
particles, it will affect only the relative motion. Therefore we
can study the problem in the coordinate system where the relative
motion is decoupled from the center-of-mass degrees of freedom.
The Hamiltonian can be rewritten (see App.~\ref{appendix:hamil})
as $H(t)=H_R(t)+H_r(t)$, where
\begin{subequations}
\begin{eqnarray}
H_R(t)&=&H_{R}^0-{\mathbf F}(t)\cdot \left( {\mathbf R}+
\frac{{\bf\bar r}_1+{\bf\bar r}_2}2\right),
\label{defHR}\\
H_r(t)&=& H_{r}^0-{\mathbf f}(t)\cdot \left( {\mathbf r+ \mathbf
d}\right)+H_1. \label{defHr}
\end{eqnarray}
\end{subequations}
Here, $H_R^0$ and $H_r^0$ contain three-dimensional harmonic
potentials, and describe also nonadiabatic effects arising when
$\omega\tau\sim 1$. In particular, $H_r^0$ incorporates terms
arising from the interaction up to the order $(a_\omega/d)^2$.
$H_1$ entails the higher-order multipole contributions.

\section{A classical model}
\label{section:class}

\begin{figure}
\epsfig{figure=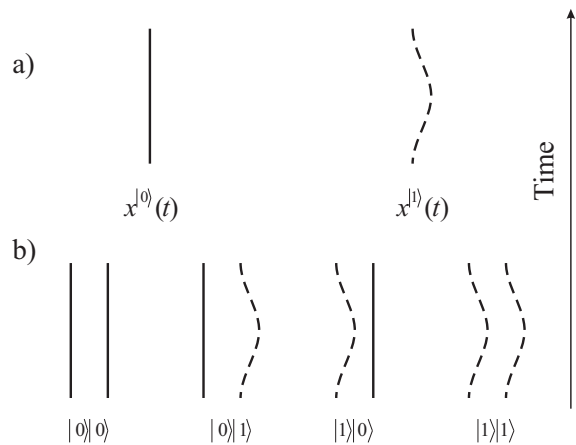,width=7.5truecm}
\caption{\label{hamilclass}Gate operation scheme in a classical
picture. Above: selective trap displacement for the two different
internal states. Below: schematics of the four different
Hamiltonians for each combination of internal states.}
\end{figure}

We first treat the ions' motion classically, i.e. we regard them
as point particles following well-defined trajectories dictated
by the state-dependent trapping potential and by the repulsive
electrostatic force. Without loss of generality, we can take the
$x$ axis parallel to $\mathbf d$. We will study the
one-dimensional problem of the motion along that direction,
denoting by italic letters the first Cartesian component of the
vectors defined in the previous Section. The initial state of the
system is described by the internal quantum state
\begin{equation}
|\chi_{\rm
cl}(t_0)\rangle\equiv\sum_{\alpha,\beta=0}^1c_{\alpha\beta}
|\alpha\rangle_1|\beta\rangle_2
\end{equation}
and by the external classical trajectories $x_i^{\alpha\beta}(t)$
of the two ions, dictated by the Hamiltonian Eq.~(\ref{defHab}).
Here, by $\alpha\rightleftharpoons\alpha_1$
($\beta\rightleftharpoons\alpha_2$) we mean the internal state of
the first (second) particle. Indeed, to find the trajectories for
all values of $\alpha$ and $\beta$, we have to solve four
distinct classical two-particle equations of motion, each
describing the dynamics for one of the possible combinations of
internal states, as depicted in Fig.~(\ref{hamilclass}).
Once we have done that, we can evaluate the Coulomb interaction
energy
\begin{eqnarray}
V^{\alpha\beta}(t)&\equiv&\frac{q_e^2}{4\pi\varepsilon_0} \frac
1{\big|d+x_2^{\alpha\beta}(t)-x_1^{\alpha\beta}(t)\big|}\nonumber\\
&=&\frac{q_e^2}{4\pi\varepsilon_0d}\sum_{n=0}^\infty
\left[\frac{x_1^{\alpha\beta}(t)-x_2^{\alpha\beta}(t)}{d}\right]^n.
\label{expVab}
\end{eqnarray}
We then define the evolved internal state as
\begin{equation}
\label{defchiclt} |\chi_{\rm cl}
(t)\rangle\equiv\sum_{\alpha\beta}c_{\alpha\beta}|\alpha\rangle|\beta\rangle
e^{i\varphi_{\rm cl}^{\alpha\beta}},
\end{equation}
where
\begin{equation}
\label{defvarphicl}
\varphi_{\rm cl}^{\alpha\beta}\equiv -\frac
1\hbar\int_{t_0}^tV^{\alpha\beta}(t')dt'.
\end{equation}
Now we make the following assumptions: (i) the force acts slowly
over the harmonic oscillator time scale $\omega^{-1}$, i.e.
$|\dot{\cal F}|\ll\omega$; (ii) it induces a displacement  of the
order of the single-trap harmonic oscillator length $a_\omega$;
(iii) the latter is much smaller than the distance between the
traps, i.e. $a_\omega\ll d$; (iv) the amplitude of the intra-well
oscillations (if any) is negligible with respect to the
inter-well distance, i.e. $E_i\ll m\omega^2d^2/2$. The first
three conditions can be fulfilled by construction, the last one
requires in principle the motion to be cooled. Assumption (i)
amounts to neglecting non-adiabatic terms in the trajectories
(e.g., sloshing motion excited by the trap displacement). On the
other hand, when the last three assumptions hold, we can
consider, to a first approximation, the intra-well motion to be
basically unaffected by the higher-order multipole terms in the
expansion of the Coulomb interaction, Eq.~(\ref{expVab}). This
approximation is not easy to check classically, since the exact
trajectories cannot be computed analytically. We will test its
validity in second-order perturbation theory, in the context of
the quantum-mechanical treatment (see
App.~\ref{appendix:statepert}).

\subsection{Starting conditions}

\begin{figure}
\epsfig{figure=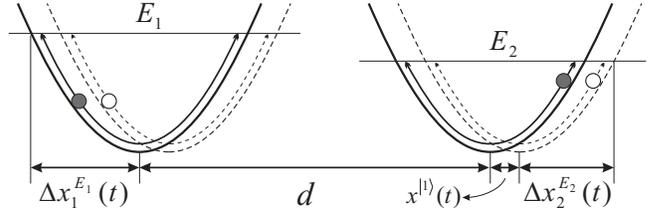,width=8.5truecm}
\caption{\label{schemeclass}Gate operation dynamics for two
classical particles oscillating with energies $E_i$. The
state-selective trap displacements $x^{\alpha_i}(t)$ and the
intra-well oscillations $\Delta x_i^{E_i}(t)$ are shown.}
\end{figure}

The initial classical motional state, at $t=t_0$, can be either
the ground state, described by the initial conditions
$x_i(t_0)=\dot x_i(t_0)=0$, or an excited state, described by
oscillations of each ion inside its trap with an energy $E_i$
($i=1,2$), i.e. by the initial conditions
\begin{subequations}
\begin{eqnarray}
\label{defDxEi}x_i(t_0)&=&\Delta x^{E_i}_i(t_0)\equiv
\sqrt{\frac{2E_i}{m\tilde\omega^2}}\cos\left[\tilde\omega(t_0-t_i)\right],\\
\dot{x_i}(t_0)&=&-\sqrt{\frac{2E_i}{m}}\sin\left[\tilde\omega(t_0-t_i)\right]
\end{eqnarray}
\end{subequations}
(of course, the former is a particular case of the latter, for
$E_1=E_2=0$). Here, $\tilde\omega$ is a corrected trap frequency,
taking into account up to quadratic terms in the Coulomb
potential --~i.e., up to $n=2$ in Eq.~(\ref{expVab})~-- namely,
$\tilde\omega\equiv\omega\sqrt{1+\epsilon/2}$, where
\begin{equation}
\epsilon\equiv \frac{q_e^2}{\pi\varepsilon_0m\omega^2d^3}
\end{equation}
is essentially twice the ratio of the Coulomb energy
$q_e^2/(4\pi\varepsilon_0d)$ and the energy of the second ion with
respect to the first trap $m\omega^2d^2/2$. Under the
approximations discussed above, we can write the trajectories
$x^{\alpha\beta}_i(t)$ as
\begin{equation}
\label{trajapprox} x^{\alpha_1\alpha_2}_i(t)\approx
x^{\alpha_i}_i(t)+\Delta x^{E_i}_i(t)
\end{equation}
at all times. The situation is depicted in
Fig.~\ref{schemeclass}.
Note that we are treating classically the particle motion, but
not the internal state: so the ions are allowed to be in a
superposition of the available logical states, i.e. to oscillate
according so to speak to two different trapping potentials, as
seen in Fig.~\ref{schemeclass}.

\subsection{Gate phases}

We can write the phases Eq.~(\ref{defvarphicl}) as
\begin{equation}
\varphi_{\rm cl}^{\alpha\beta}=\phi_{\rm
cl}^{\alpha\beta}+\delta\phi^{\alpha\beta},
\end{equation}
where $\phi_{\rm cl}^{\alpha\beta}$ is the ground-state
contribution, and $\delta\phi^{\alpha\beta}$ is the correction
due to motional excitations. To evaluate the various
contributions explicitly under the above approximations, we now
need only to specify the time dependence of the trap
displacement. We choose the Gaussian form
\begin{equation}
\label{defcalF} {\cal F}(t)\equiv\xi e^{-(t/\tau )^2}.
\end{equation}
Let us first consider the case where both particles are in their
motional ground state. We insert Eq.~(\ref{trajapprox}) for
$E_i=0$, through Eq.~(\ref{expVab}), into
Eq.~(\ref{defvarphicl}), and obtain
\begin{equation}
\label{evalphicl} \phi_{\rm
cl}^{\alpha\beta}=-(\alpha-\beta)^2\sqrt{\frac\pi
8}\xi^2\epsilon\tilde\omega\tau\sum_{n=0}^\infty
\frac{[(\alpha-\beta)\xi
a_{\tilde\omega}/d]^{n-1}}{\sqrt{2(n+1)}}.
\end{equation}
In the evaluation of the ground-state phase $\phi_{\rm
cl}^{\alpha\beta}$, Eq.~(\ref{evalphicl}), the complete Coulomb
potential Eq.~(\ref{expVab}) has been taken into account. When we
evaluate the corrections $\delta\phi^{\alpha\beta}$ instead, it
is not possible to find a general expression valid at all orders
$n$, which therefore have to be considered separately. We choose
$t_0$ as an integer multiple of the oscillation period
$2\pi/\tilde\omega$ (so that the motional state is left unchanged
after gate operation), and find
\begin{eqnarray}
\label{evaldphith} \delta\phi^{\alpha\beta}&=& 3 (\alpha-\beta)
\sqrt{\frac\pi
8}\xi^2\epsilon\tilde\omega\tau\!\left[\frac{1}{\sqrt
2\xi}\frac{a_{\tilde\omega}}d
+(\alpha-\beta)\left(\frac{a_{\tilde\omega}}d\right)^2\right]\nonumber\\
&&\mbox{}\times\frac 1{\hbar\tilde\omega}
\Big\{E_1+E_2+2\sqrt{E_1E_2}\cos[\tilde\omega(t_1-t_2)]\Big\}\nonumber\\
&&\mbox{}+o[(a_{\tilde\omega}/d)^3].
\end{eqnarray}
where it has been taken into account that $\omega\tau\gg 1$. The
two terms under square brackets in Eq.~(\ref{evaldphith}) come
from terms in the Coulomb potential with $n=3$ and $n=4$ in
Eq.~(\ref{expVab}), respectively. This means that no thermal
correction is to be expected if only harmonic contributions to
the potential ({\em i.e.}, with $n\leq 2$) are included. Indeed,
in this case the spurious interaction phases, due to the
oscillations in the ions' positions, are averaged out when
integrating on a time much larger than the oscillation period.
This explains intuitively why the phase does not depend on the
motional state, in the approximation where only linear and
quadratic terms in the Coulomb potential are taken into account,
as it will be shown analytically in Sec.~\ref{section:quadr}. Now,
the classical analogue of the gate phase Eq.~(\ref{defvartheta})
can be written as
\begin{eqnarray}
\label{varthetaclass} \vartheta_{\rm
cl}&\equiv&\sum_{\alpha,\beta}(-1)^{\alpha+\beta}
\varphi^{\alpha\beta}_{\rm cl}\nonumber\\
&=&\sqrt\frac\pi 8 \frac{\epsilon\omega\tau}{(a_\omega/d)^2}\,{\rm
Li}_{1/2}\left[(\xi a_\omega/d)^2\right]\nonumber\\
&=&\theta_{\rm cl}+o[(a_{\tilde\omega}/d)^2],
\end{eqnarray}
where ${\rm Li}_{n}(z)\equiv\sum_{k=1}^\infty z^k/k^n$ is the {\em
polylogarithm function}, and
\begin{equation}
\label{thetaclass}\theta_{\rm cl}\equiv\sqrt{\frac\pi
8}\xi^2\epsilon\tilde\omega\tau.
\end{equation}
From Eqs.~(\ref{varthetaclass}) and (\ref{thetaclass}), we see
that in our one-dimensional classical model, up to first order in
powers of $a_{\tilde\omega}/d$, the gate phase is insensitive to
motional excitations inside each trap. Later on (see
Sec.~\ref{section:many}), we will find that the very same
expression for the gate phase can be obtained, under the same
approximations, with the full three-dimensional quantum formalism.
Moreover, note that assumption (ii) applied to Eq.~(\ref{defcalF})
means $\xi\approx 1$, whence (i) implies $\omega\tau\gg 1$. It
follows that, if we want to obtain $\theta_{\rm cl}=\pi$, it has
to be $\epsilon\ll 1$. This means that the confinement has to be
strong with respect to Coulomb interaction over the inter-well
separation, which is in turn consistent with assumption (iii).

We will now consider a more general initial condition than the
ones discussed so far, namely a thermal state, described by a
probability distribution over the energies $E_i$ and the
oscillation phases $\tilde\omega t_i$. Assuming the energy
distribution characteristic of a canonical ensemble and a uniform
probability distribution for $\omega t_1$ and $\omega t_2$, we can
compute the thermal averaged phase
\begin{eqnarray}
\langle\!\langle\varphi_{\rm
cl}^{\alpha\beta}\rangle\!\rangle&\equiv&
\int_0^{2\pi/\omega}\!\frac{dt_1dt_2}{(2\pi/\omega)^2}\int_0^\infty
\frac{dE_1dE_2}{(k_BT)^2}\;
\varphi_{\rm cl}^{\alpha\beta}e^{-\frac{E_1+E_2}{k_BT}}\nonumber\\
&=&-\theta_{\rm cl}\left\{\frac 12+\frac{\alpha-\beta}{\sqrt
2\xi}\left[\frac d{a_{\tilde\omega}}+
\frac{a_{\tilde\omega}}d\left(\frac{\xi^2}{\sqrt
3}+\frac{6k_BT}{\hbar\tilde\omega}\right)\right]\right.\nonumber\\
&&\left.\mbox{}+\left(\frac{a_{\tilde\omega}}d\right)^2\left(\frac{\xi^2}{2\sqrt
2}+\frac{6k_BT}{\hbar\tilde\omega}\right)\right\}\delta_{\alpha\beta}
+o[(a_{\tilde\omega}/d)^3].\nonumber\\
\end{eqnarray}
The mean gate phase turns out to be
\begin{eqnarray}
\label{meanvarthetacl} \langle\!\langle\vartheta_{\rm
cl}\rangle\!\rangle&\equiv&\sum_{\alpha,\beta}(-1)^{\alpha+\beta}
\langle\!\langle\varphi^{\alpha\beta}_{\rm cl}\rangle\!\rangle\\
&=&\theta_{\rm
cl}\left[1+\left(\frac{a_{\tilde\omega}}d\right)^2\left(\frac{\xi^2}{\sqrt
2}+\frac{6k_BT}{\hbar\tilde\omega}\right)\right]+o[(a_{\tilde\omega}/d)^3],\nonumber
\end{eqnarray}
Indeed, we will see that the very same structure for the
corrections to the lowest-order phase is obtained with the full
quantum-mechanical calculation.

\subsection{Gate fidelity}

\begin{figure}
\epsfig{figure=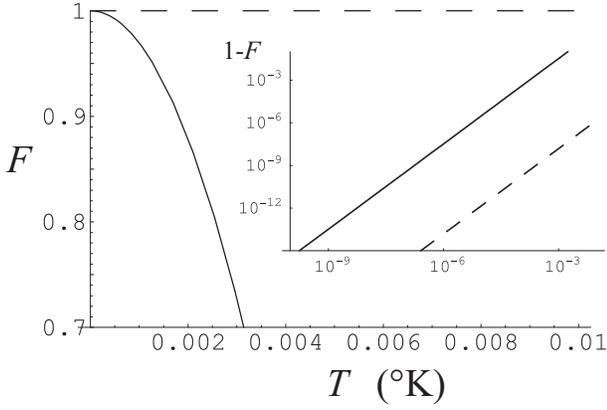,width=8truecm}
\caption{\label{figure:fidelity} Fidelity $F_{\rm cl}$ (solid
line) and improved fidelity $F'_{\rm cl}$ (dashed line) in the
classical model as a function of temperature $T$. Inset: detail
of the departure from unity of the same quantities, on a
logarithmic scale. We assumed to work with Ca$^+$ ions and chose
the parameters $\omega=2\pi\times 1{\rm MHz}$, $d=20\mu{\rm m}$.}
\end{figure}

In order to obtain the desired phase gate Eq.~(\ref{defGate}), we
require that $\langle\!\langle\vartheta_{\rm
cl}\rangle\!\rangle=\pi$. So the reference state, representing
the ideal evolution, is chosen as
\begin{equation}
\label{defchiclp} |\chi_{\rm
cl}'\rangle\equiv\sum_{\alpha\beta}c_{\alpha\beta}
|\alpha\rangle|\beta\rangle
e^{i\langle\!\langle\varphi^{\alpha\beta}_{\rm
cl}\rangle\!\rangle}.
\end{equation}
The real evolved state, Eq.~(\ref{defchiclt}), can be written as
$|\chi_{\rm cl}(t)\rangle=|\chi_{\rm cl}'\rangle+|\delta\chi_{\rm
cl}\rangle$, whereby
\begin{equation}
\label{defdeltachi} |\delta\chi_{\rm
cl}\rangle\equiv\sum_{\alpha\beta}c_{\alpha\beta}
|\alpha\rangle|\beta\rangle\left( e^{i\varphi^{\alpha\beta}_{\rm
cl}}-e^{i\langle\!\langle\varphi^{\alpha\beta}_{\rm
cl}\rangle\!\rangle}\right)
\end{equation}
In our classical model, we are treating our particles' external
degrees of freedom classically. Therefore, in the evaluation of
the fidelity Eq.~(\ref{defFid}), instead of tracing over the
motional eigenstates we should average over the possible
classical trajectories. Thus
\begin{eqnarray}
\label{defFcl} F_{\rm cl}&\equiv&
\min_\chi\Big\langle\!\!\Big\langle \langle\chi_{\rm
cl}'|\chi_{\rm cl}(t)\rangle\langle\chi_{\rm cl}(t)|\chi_{\rm
cl}'\rangle\Big\rangle\!\!\Big\rangle\nonumber\\
&=&\min_\chi\Big\langle\!\!\Big\langle \left|1+\langle\chi_{\rm
cl}(t)|\delta\chi_{\rm
cl}\rangle\right|^2\Big\rangle\!\!\Big\rangle\nonumber\\
&\stackrel{(\ref{defdeltachi})}=&
\min_{\{c_{\alpha\beta}\}}\Big\langle\!\!\Big\langle
\bigg|\sum_{\alpha,\beta=0}^1\left|c_{\alpha\beta}\right|^2
e^{-i\left(\varphi^{\alpha\beta}_{\rm
cl}-\langle\!\langle\varphi^{\alpha\beta}_{\rm
cl}\rangle\!\rangle\right)}\bigg|^2\Big\rangle\!\!\Big\rangle\nonumber\\
&=&1-\left(\frac{6\theta_{\rm cl}
k_BT}{\hbar\tilde\omega}\right)^2\left[\frac 1{\xi^2}
\left(\frac{a_{\tilde\omega}}d\right)^2-2\left(\frac{a_{\tilde\omega}}d\right)^4\right]\nonumber\\
&&\mbox{}+o[(a_{\tilde\omega}/d)^5],\label{evalFidcl}
\end{eqnarray}
as is discussed in detail in App.~\ref{appendix:fidclass}.
Finally, let us consider what would come out if we were able to
suppress the cubic anharmonic correction from the Coulomb
potential, i.e. to put $\kappa=0$. We will show later
(Sect.~\ref{section:qfid}) how this can be done in practice --
here we would like to give a classical estimate $F'_{\rm cl}$ of
the improved gate fidelity. The calculation is performed in
App.~\ref{appendix:fidclass} as well, and the result is
\begin{equation}
\label{evalFcl4} F'_{\rm cl}(T)=1-\left(\frac{3\theta_{\rm cl}
k_BT}{\hbar\tilde\omega}\right)^2
\left(\frac{a_{\tilde\omega}}d\right)^4+o[(a_{\tilde\omega}/d)^5],
\end{equation}
This shows that, by suppressing one order of anharmonic
corrections, one obtains an improvement by two orders in
$a_\omega/d$ (several orders of magnitude) in the fidelity, as is
shown in Fig.~\ref{figure:fidelity}.

\section{Quantum treatment}
\label{section:quantum}

\begin{figure}
\epsfig{figure=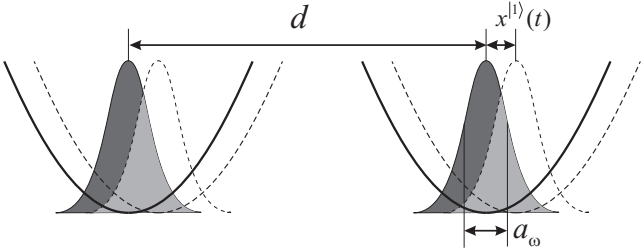,width=8.5truecm}
\caption{\label{schemequant}Gate operation scheme in the quantum
regime. Trap parameters are defined like in the classical case; the
harmonic-oscillator ground-state width $a_\omega$ is also shown.}
\end{figure}

We want to describe quantum mechanically the three-dimensional dynamics of the
two particles. This means that, unlike in the previous Section, their motional
state is given by a wavefunction (see Fig.~\ref{schemequant})
which evolves according to the Hamiltonian Eq.~(\ref{defHab}).
To better understand its structure, it is useful to write
\begin{equation}
H^{\alpha\beta}(t,\mathbf{\hat x}_1,\mathbf{\hat x}_2)\equiv
H^{\alpha\beta}\big(t,\mathbf{x}^\alpha(t),\mathbf{x}^\beta(t)\big)+
H^{\alpha\beta}_e(t,\mathbf{\hat x}_1,\mathbf{\hat x}_2).
\label{rewrHab}
\end{equation}
Here the $\mathbf{\hat x}_j$ are the ion position operators, and
the c-numbers $\mathbf{x}^{\alpha_j}(t)$ denote the centers of
the initial motional wavefunctions as determined by the trap (see
Sect.~\ref{section:cond}). To second order in the expansion
Eq.~(\ref{expVab}), the first term at the right-hand side of
Eq.~(\ref{rewrHab}) gives rise to the same contribution to the
phase already calculated in lowest order in the classical model,
namely $\theta_{\rm cl}$ given in Eq.~(\ref{thetaclass}).
Corrections to this phase are due, as before, to: (a) thermal
excitations; (b) higher-order (multipole) terms in the expansion
of the Coulomb potential; (c) nonadiabaticity. Motional effects
of the kinds (a) and (c) are accounted for by
$H^{\alpha\beta}_e$, while multipole corrections get a
contribution also from
$H^{\alpha\beta}\big(t,\mathbf{x}^\alpha(t),\mathbf{x}^\beta(t)\big)$.
In order to minimize such corrections, we choose to operate in
the adiabatic regime, given by condition (i) in
Sect.~\ref{section:class}, i.e. we assume $\omega\tau\gg 1$. We
study the dynamics in the center-of-mass and relative coordinate
system, as given by Eqs.~(\ref{defHR}) and (\ref{defHr}). In both
coordinate systems, the motion along different axes decouples:
the transverse directions contribute just an overall phase,
whereas the relevant state-dependent dynamics takes place along
the $x$ axis. Since we assumed $d\gg a_\omega$, it follows that
$H_1$ --~containing only terms of $o[(a_\omega/d)^3]$~-- can be
treated as a small perturbation in $H$. We will first neglect it
and solve exactly the three-dimensional Schr\"{o}dinger equation,
and then take it into account perturbatively, eventually checking
our results with a numerical simulation.

\subsection{Unperturbed forced oscillator}
\label{section:quadr}

When we put $H_1=0$ and take ${\cal F}(t)$ as in Eq.~(\ref{defcalF}),
the solution to Eq.~(\ref{Schroed}) can be written
explicitly \cite{Galindo}. This is done in App.~\ref{appendix:unpert}. The
gate phase turns out to be, in this approximation,
\begin{equation}
\label{deftheta}
\theta\equiv\sum_{\alpha,\beta}(-1)^{\alpha+\beta}\phi^{\alpha\beta}=
2\xi ^2\left[ \Phi ( \omega) -\Phi(\omega\sqrt{1+\epsilon}) \right],
\end{equation}
where
\begin{subequations}
\begin{eqnarray}
\Phi(\omega)&\equiv&
-\Im\left[\int_{t_0}^tds\,K(s,t_0)\frac{dK^{*}(s,t_0)}{ds}\right],\\
K(t,t_0) &\equiv &\frac 1{\sqrt{m\hbar\omega}}
\int_{t_0}^tdt'\,{\cal F}(t')e^{i\omega (t'-t_0)}.
\end{eqnarray}
\end{subequations}
Explicit expressions for $\phi^{\alpha\beta}$ and for $\Phi(\omega)$ are given
in Eqs.~(\ref{phiab})-(\ref{Phiab}). In the limit $\omega\tau\gg 1$, we obtain
$\Phi(\omega)\approx-\sqrt{\pi/32}\,\omega\tau$. By expanding
Eq.~(\ref{deftheta}) up to first order in $\epsilon$, we retrieve
$\theta\approx\theta_{\rm cl}$ as given by Eq.~(\ref{thetaclass}).
The phase $\theta$ can be adjusted to
the desired value $\pi$ by tuning the displacement $\xi$ and/or the
interaction time $\tau$.
Moreover, $\theta$ is independent of the ions' motional state. This means that
the phase remains the same even
if we start with a mixed external state, described by a
density matrix
\begin{equation}
\label{defrhoT} \rho_T(t_0)\equiv \frac{e^{-H(t_0)/k_BT}}{Z}
\approx (1-\gamma)^6\bigotimes_{i=1}^6\sum_{n_i=1}^\infty
\gamma^{n_i}|n_i\rangle_i\langle n_i|,
\end{equation}
corresponding to a thermal distribution at a temperature $T$.
Here the canonical partition function $Z\equiv{\rm
tr}\left\{e^{-H(t_0)/k_BT}\right\}$,
$\gamma\equiv\exp(-\hbar\omega/k_BT)$,
$\{n_i\}_{i=1,\cdots,6}\equiv\{n_x,n_y,n_z,N_X,N_Y,N_Z\}$, and
the $\left|N\right\rangle _{X,Y,Z}$ ($\left| n\right\rangle
_{x,y,z}$) are the eigenstates of $H_{{R}}^0$ ($H_{r}^0$) along
each direction. To optimize the gate fidelity (see
Sect.~\ref{section:defgate}), we need basically one thing more --
that the external degrees of freedom are not entangled with the
internal ones after gate operation, i.e. that the final motional
state does not depend on the logical states of the qubits. This
indeed happens, under the adiabatic assumption (i) of
Sect.~\ref{section:class}. In fact, in this case, the overlap
${\cal O}^{\{n_i\}}(t,t_0)$ between the initial and final spatial
wavefunctions for a system starting in a motional eigenstate
along all degrees of freedom, defined by Eq.~(\ref{OROr}), is
close to 1. To be more precise, ${\cal O}^{\{n_i\}}(t,t_0)$
formally depends on the motional state -- however, if the
adiabatic condition is satisfied and $|t|$, $|t_0|$ are large
enough, ${\cal O}^{\{n_i\}}(t,t_0)$ tends exponentially to 1, as
it can be seen from Eqs.~(\ref{defOR})-(\ref{defOr}).

\subsection{Including higher-order terms}
\label{section:corr}

We will now take into account the contribution of $H_1$.
This does not affect the center-of-mass motion.
Therefore from now on we will study only the relative motion. A little care is
needed since,
unlike in the most common time-dependent perturbation theory, in our case
the unperturbed Hamiltonian depends on time, while the perturbation does not.
The calculation is carried out in App.~\ref{appendix:pert}. In the adiabatic
limit and for $|t|,|t_0|>\tau$, first-order corrections simply amount
to a relatively small additional phase shift:
\begin{equation}
\langle U(t,t_0)\rangle\approx \langle U_0(t,t_0)\rangle e^{i
(\Delta^{\alpha\beta}+\Delta')}.\label{Upert1}
\end{equation}
Thus, to this order,
\begin{equation}
\label{varphipert}
\varphi^{\alpha\beta}\approx\phi^{\alpha\beta}+\Delta^{\alpha\beta}+\Delta'.
\end{equation}
Hence we find $\vartheta=\theta+\delta\theta+o[(a_\omega/d)^6]$,
where
\begin{equation}
\label{defdtheta} \delta\theta=8\frac{a_\omega^2}{d^2}\theta_{\rm
cl}\! \left[\sqrt{2}\xi^2 +3(2n_x-n_y-n_z)\right],
\end{equation}
and it is understood that we started from a pure state with
$n_{x,y,z}$ excitations along the various directions of the
relative motion. As already anticipated,
Eq.~(\ref{defdtheta}) has the same structure as the classical
correction to the phase expressed by Eq.~(\ref{meanvarthetacl}),
except for a different overall factor related to the
dimensionality of the problem we are now considering. Indeed, the
first term under square brackets comes from the higher multipoles
($k=3$) in $H_1$, while the second one is due to the
thermal excitations.

\subsection{Numerical computation}
\label{section:num}

\begin{figure}
\epsfig{figure=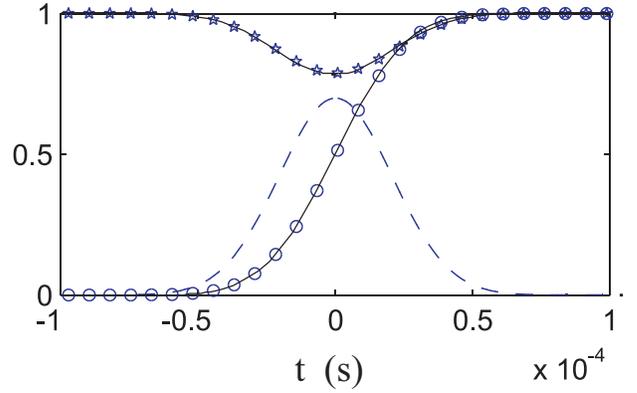,width=8.5truecm} \caption{\label{phaseq}Gate phase
$\vartheta/\pi$ (circles), and projection of the initial motional
ground state over the evolved one for ions in states $|0\rangle_1|1\rangle_2$
(stars), as induced by the external force ${\cal
F}(t)$ (dashed line). The results of the numerical calculation,
performed with the parameters quoted in the text including corrections up to
$o[(a_\omega)^5]$, are shown. Solid lines are analytical results in the
harmonic approximation of Sect.~\protect\ref{section:quadr}.}
\end{figure}

In order to check the validity of the perturbative expression
Eq.~(\ref{defdtheta}), we solved numerically the Schr\"odinger
equation for the relative motion, taking into account cubic and
quartic interaction terms, explicitly given by Eqs.~(\ref{defP3})
and (\ref{defP4}) respectively. The calculation is described in
App.~\ref{appendix:num}, and results are shown in Fig.~\ref{phaseq}.
In particular we find that the cubic
corrections cancel indeed each other, and that the quartic
corrections have the same order of magnitude as predicted by
Eq.~(\ref{defdtheta}). From Eq.~(\ref{deftheta}) we obtain
$\theta=\pi$ with $\omega=2\pi\times 1$MHz, $\bar x_2-\bar
x_1=20\mu$m, $\xi=0.7$ and $\tau=41.1069\mu$s. These results
were confirmed by the numerical computation up to 40 initial
excitations in each direction, always giving unity overlap of the
final motional state on the initial one. Indeed, with
$t=-t_0=150\mu$s, even starting e.g. with the $(10^{4})^{\rm th}$
oscillator excited state, Eq.~(\ref{OROr}) still predicts ${\cal
O}(t,t_0)>1-10^{-10}$. With these parameters, the perturbative
estimate derived within the classical model in
Sec.~\ref{section:class} turns out to be $\theta_{\rm cl}\approx
1.04\pi$.

\subsection{Gate fidelity}
\label{section:qfid}

The gate phase $\vartheta$ cannot be measured directly, since the higher-order
corrections arising from the Coulomb potential depend both on the internal and
on the motional state of each ion, and cannot be undone by means of
single-qubit operations, unless the logical state is measured. However, the
corrections of order $k$ have a simple internal-state dependence of the kind
$(\alpha-\beta)^k$, as shown by Eq.~(\ref{defDeltaab}).
This implies that it is possible to obtain a cancellation of the odd-order
corrections, by applying a $\pi$-pulse
$R\equiv|0\rangle\langle 1|+|1\rangle\langle 0|$ to both qubits in the middle
of gate operation. Indeed, if $U$ is the evolution operator giving the dynamics
described in the previous Sections, we find
\begin{equation}
\begin{array}{rcrcrl}
|0\rangle|0\rangle &\stackrel{RU}\longrightarrow
&e^{i\varphi^{00}}|1\rangle|1\rangle
&\stackrel{RU}\longrightarrow
&e^{i(\varphi^{00}+\varphi^{11})}|0\rangle|0\rangle&\\
|0\rangle|1\rangle &\longrightarrow &e^{i\varphi^{01}}|1\rangle|0\rangle
&\longrightarrow &e^{i(\varphi^{01}+\varphi^{10})}|0\rangle|1\rangle&\\
|1\rangle|0\rangle &\longrightarrow &e^{i\varphi^{10}}|0\rangle|1\rangle
&\longrightarrow &e^{i(\varphi^{10}+\varphi^{01})}|1\rangle|0\rangle&\\
|1\rangle|1\rangle &\longrightarrow &e^{i\varphi^{11}}|0\rangle|0\rangle
&\longrightarrow &e^{i(\varphi^{11}+\varphi^{00})}|1\rangle|1\rangle&\!\!\!.
\end{array}
\end{equation}
Here an adiabatic approximation is understood, according to which the final and
initial motional state are identical. We now define
\begin{equation}
\Delta\theta\equiv\delta\theta-\langle\!\langle\delta\theta\rangle\!\rangle
\end{equation}
(as before, $\langle\!\langle\cdot\rangle\!\rangle$ denotes the thermal
average), and the gate operator
\begin{equation}
G\equiv S(RU)^2,
\end{equation}
where
\begin{equation}
\label{defS}
S\equiv|0\rangle\langle 0|e^{-2i\xi^2\Phi(\omega)}+
|1\rangle\langle 1|e^{-i[2\xi^2\Phi(\omega\sqrt{1+\epsilon})-
\langle\!\langle\delta\theta\rangle\!\rangle]}.
\end{equation}
If we choose the gate operation time $\tau$ in such a way that
\begin{equation}
\pi\stackrel!=4\xi ^2\left[ \Phi ( \omega)
-\Phi(\omega\sqrt{1+\epsilon})
\right]-2\langle\!\langle\delta\theta\rangle\!\rangle,
\end{equation}
we obtain
\begin{equation}
\begin{array}{rcrl}
|0\rangle|0\rangle &\stackrel{G}
\longrightarrow &e^{i\Theta}|0\rangle|0\rangle&\\
|0\rangle|1\rangle &\longrightarrow
&e^{-i\Delta\theta}e^{i\Theta}|0\rangle|1\rangle&\\
|1\rangle|0\rangle &\longrightarrow
&e^{-i\Delta\theta}e^{i\Theta}|1\rangle|0\rangle&\\
|1\rangle|1\rangle &\longrightarrow
&e^{i\pi}e^{i\Theta}|1\rangle|1\rangle&\!\!\!,
\end{array}
\label{defgateG}
\end{equation}
The global (thus irrelevant) phase $\Theta$ is given in
App.~\ref{appendix:fid}. Note that the single-qubit rotation $S$
is the same for both qubits, and therefore single-ion
addressability is not required. The fidelity of the gate
operation is defined by comparing the ideal gate operation
Eq.~(\ref{defGate}) with the actual dynamics obtained in our
scheme, Eq.~(\ref{defgateG}), at a given temperature $T$ for the
motion in all three dimensions. The result, derived in
App.~\ref{appendix:fid}, is
\begin{equation}
F(T)\approx1- 6^3\left(\frac{\theta_{\rm cl}k_BT}
{\hbar\omega}\right)^2\left(\frac{a_\omega}d\right)^4.
\label{evalF}
\end{equation}
The fidelity turns out to be independent of $\tau$ and $\xi$
which, subject to the conditions $\omega\tau\gg 1$ and $\xi\sim
1$, can be freely chosen to obtain the desired gate phase. The
dependence of the fidelity on the various parameters is the same
as in the classical model discussed in the previous Section. As
already anticipated in the previous Section, the intermediate
$\pi$-pulse $R$ allows us to get rid of the $o[(a_\omega/d)^2]$
term, thus obtaining a much better gate performance. Indeed, the
only difference between the corrected classical fidelity $F'_{\rm
cl}$ and the quantum fidelity $F$ is the numerical pre-factor
multiplying the temperature-dependent part, which is bigger in
the latter case due to the inclusion of all the spatial degrees
of freedom, whereas our classical model was just one-dimensional.
Anyway, with the parameters quoted above, at temperatures below 2
mK, corresponding to an average number of harmonic-oscillator
excitations $\bar n\sim 6$, the fidelity turns out to be bigger
than $1-10^{-6}$. We can also evaluate how the fidelity scales
when the gate is repeatedly applied, say $g$ times. It is clear
from Eq.~\ref{defgateG} that in this case, apart again from an
overall phase,
\begin{equation}
\begin{array}{rcrl}
|0\rangle|0\rangle &\stackrel{G^g}
\longrightarrow &|0\rangle|0\rangle&\\
|0\rangle|1\rangle &\longrightarrow
&e^{-ig\Delta\theta}|0\rangle|1\rangle&\\
|1\rangle|0\rangle &\longrightarrow
&e^{-ig\Delta\theta}|1\rangle|0\rangle&\\
|1\rangle|1\rangle &\longrightarrow
&(-1)^g|1\rangle|1\rangle&\!\!\!.
\end{array}
\label{nGate}
\end{equation}
The excitation-dependent phase $\Delta\theta$ is just multiplied by $g$.
Thus, under the
same approximations as above, the fidelity of the $g$-fold gate operation is
\begin{equation}
F^{(g)}(T)=1-g^2[1-F(T)],
\end{equation}
i.e. it scales with the square of the number of gates.

\subsection{One-dimensional calculation for many ions}
\label{section:many}

We now assume to have $N$ ions, trapped in a linear array of equally spaced
traps, i.e. we take
\begin{equation}
{\bf\bar r}_j\equiv(\bar x_j,\bar y_j,\bar z_j)=j {\bf d}
\end{equation}
in Eq.~(\ref{defHi}).
Expanding the interaction Hamiltonian $H_{ij}$, Eq.~(\ref{defHij}), in powers of
the ${\bf r}_i$ and ${\bf r}_j$, and neglecting terms of
$o[(a_\omega/d)^3]$, we find
\begin{eqnarray}
H&\approx &\frac{m\omega^2}2\Bigg\{\sum_{i=1}^N\bigg[\frac{\omega_i^2}{\omega^2}
(x_i-\tilde x_i)^2+y_i^2+z_i^2-\varepsilon_i\nonumber\\
&&\mbox{}-2a_\omega{\cal F}(t)|1\rangle_i\langle 1| (x_i+\bar
x_i)\bigg]+\epsilon\sum_{i<j}^N\frac{x_ix_j}{|i-j|^3}\Bigg\},\qquad
\label{HNions}
\end{eqnarray}
where the various quantities are defined in App.~\ref{appendix:many}.
Eq.~(\ref{HNions}) describes a set of independent forced harmonic oscillators,
like the ones we solve in App.~\ref{appendix:unpert}, plus a
coupling term multiplied by $\epsilon$. If $\epsilon\ll 1$, we can treat this
term as a small perturbation, in the very same way we develope in
App.~\ref{appendix:pert}. We take as initial state
\begin{equation}
|\Psi_N(t_0)\rangle=\prod_{i=1}^N|{\bf n}_i\rangle_i|\alpha_i\rangle_i,
\end{equation}
where $\alpha_i$ denotes the internal state of the $i^{\rm th}$ ion, and
${\bf n}_i$ its motional state in the well corresponding to the $i^{\rm th}$
term of the first sum at the right-hand side of Eq.~(\ref{HNions}).
We obtain
\begin{equation}
|\Psi_N(t)\rangle\approx\frac 12\prod_{i=1}^N\bigg(e^{i\phi^{\alpha_i}}
\prod_{j\not=i}e^{i\phi^{\alpha_i\alpha_j}}\bigg)|\Psi_N(t_0)\rangle,
\end{equation}
where (calculating the two-particle phases $\phi^{\alpha_i\alpha_j}$
perturbatively)
\begin{subequations}
\begin{eqnarray}
\phi^{\alpha_i}&\approx &\alpha_i^2\xi^2\Phi(\omega_i)
+\alpha_i\sqrt{\pi}\omega\tau\xi\bar x_i/a_\omega\nonumber\\
&&\mbox{}-[n_{x,i}\omega_i+(n_{y,i}+n_{z,i})\omega](t-t_0),\\
\phi^{\alpha_i\alpha_j}&\approx &
\epsilon\int_{t_0}^t\!\!\langle\Psi_N(t_0)|U_0(t,t')\frac{x_ix_j}{|i-j|^3}
U_0(t',t_0)|\Psi_N(t_0)\rangle dt'\nonumber\\
&=&\sqrt{\frac\pi 8}\frac{\alpha_i\alpha_j}{|i-j|^3}
\frac{\xi^2\epsilon\omega\tau}{(1+\epsilon\eta_i)(1+\epsilon\eta_j)}\nonumber\\
&\approx &\frac{\alpha_i\alpha_j}{|i-j|^3}\theta_{\rm cl},\label{phiaiaj}
\end{eqnarray}
the last line following from $\epsilon\ll 1$ and Eq.~(\ref{maxeta}). Again, the
result to this order turns out to be independent of the motional state of anyone
of the ions. In the case of two ions, Eq.~(\ref{phiaiaj}) gives back
Eq.~(\ref{thetaclass}).
\end{subequations}

\section{Conclusions}

We analyzed in detail a recent proposal \cite{ion2000} for
scalable quantum computation with ions in an array of microtraps.
This scheme bears important advantages over the previous proposal
\cite{ion95}, based on trapped ions as well. Indeed, in that case
many ions, lying in a single trap minimum, exchange information
via the collective motional excitations; ground-state cooling is
an absolute need, and any perturbation on each ion can affect the
performance of the whole system. Here, instead, each ion is
confined to a single minimum of a periodic microscopic potential,
and interacts with other ions via Coulomb force. Under the
conditions discussed in the text (adiabaticity of the trap
displacement, strong confinement with respect to the distance
between ions, intermediate symmetrizing $\pi$-pulse), the phase
shift is insensitive, to a high accuracy, on the motional state
of each ion inside each trap, and therefore the fidelity turns
out to be practically independent from temperature. Moreover,
trapping frequencies can be much higher than in the previous
case, leading to much shorter gate operation times. As long as we
take into account purely motional decoherence mechanisms, we find
a fidelity bigger than $1-10^{-6}$ for a two-qubit phase gate
operating on a time scale of a few tens of $\mu$s. Furthermore,
with the improved scheme presented here, single-qubit
addressability is not required for any of the various control
operations. To sum up, the present proposal constitutes really a
good candidate for a scalable implementation of a quantum
computer.

\acknowledgments

This research has been supported by the Austrian Science
Foundation, the Institute for Quantum Information GmbH, and the
European Commission through contracts ERB-FMRX-CT96-0087 and
HPMF-CT-1999-00211. T.C. acknowledges support from Istituto
Trentino di Cultura.

\begin{appendix}

\section{Classical calculation}

In this Appendix we give a detailed account of the calculations
leading to the results obtained in the classical model for our
two-qubit phase gate.

\subsection{Rewriting the Hamiltonian}
\label{appendix:hamil}

In this Section we give the explicit form of the various terms in
Eqs.~(\ref{defHR}) and (\ref{defHr}), describing the Hamiltonian
Eq.~(\ref{HamN}) for two ions, in the center-of-mass and
relative-motion coordinate systems. Indeed the Hamilonian, with
the number of ions $N=2$, may be rewritten as $H=H_{R}+H_{r}$,
where
\begin{subequations}
\begin{eqnarray}
H_{R} &=&\frac{{\mathbf P}^2}{2M}+\frac 12M\omega ^2{\mathbf R}^2-
{\mathbf F}(t)\cdot \left( {\mathbf R+\mathbf R}_0\right)\\
H_{r} &=&\frac{{\mathbf p}^2}{2\mu }+\frac 12\mu \omega
^2({\mathbf r}+\mathbf d-\mathbf d_0)^2\nonumber\\
&&\qquad\mbox{}-{\mathbf f}(t)\cdot \left( {\mathbf r+ \mathbf
d}\right) +\frac \lambda {|{\mathbf r+\mathbf d}|
}\quad\label{defHrApp}
\end{eqnarray}
and
\end{subequations}
\begin{subequations}
\begin{eqnarray}
{\mathbf R} &\equiv &\left( X,Y,Z\right) \equiv \frac{{\mathbf r}_1+
{\mathbf r}_2}2-{\mathbf R}_0,\\
\mathbf r&\equiv &\left( x,y,z\right) \equiv {\mathbf r
}_2-{\mathbf r}_1-{\mathbf  d},
\end{eqnarray}
\begin{align}
{\mathbf R}_0 &\equiv \frac{\bar{\mathbf r}_1+ \bar{\mathbf
r}_2}2,&\mathbf d_0&\equiv|\bar{\mathbf r}_2-\bar{\mathbf r}_1|
=(d_0,0,0),\nonumber\\
\label{defd0}\\
{\mathbf P} &\equiv {\mathbf p}_1+{\mathbf p}_2,&{\mathbf
p}&\equiv \frac{{\mathbf p}_2-{\mathbf p}_1}2,\\
{\mathbf F}(t) &\equiv {\mathbf F}_1(t)+{\mathbf F}_2(t),
&{\mathbf f}
(t)&\equiv \frac{{\mathbf F}_2(t)-{\mathbf F}_1(t)}2,\\
M &\equiv 2m, \quad\mu \equiv \frac m2, &\lambda &\equiv
\frac{q_e^2}{ 4\pi \varepsilon _0}.
\end{align}
In the above, $\mathbf d\equiv(d,0,0)$ is the equilibrium
separation between the two particles in the absence of the
pushing force. Due to the repulsive Coulomb interaction,
$|\mathbf d|\equiv d$ will be bigger than the distance $d_0$
between the centers of the two bare harmonic traps, defined in
Eq.~(\ref{defd0}). The correction $\delta x\equiv d-d_0$ is the
solution of the equation
\end{subequations}
\begin{equation}
0=\frac\partial{\partial x}\left(\frac 12\mu \omega ^2x^2 +\frac
\lambda {|x+d_0|}\right)=
\mu\omega^2x-\frac\lambda{(x+d_0)^2}\label{CoulMin}
\end{equation}
(it has been assumed $d_0+x>0$), and can be written as
\begin{equation}
\delta x=\frac {4d_0}3 \sinh^2\left\{\frac
16\ln\left[\eta+1+\sqrt{\eta(\eta+2)}\right]\right\},
\end{equation}
with $\eta\equiv\lambda/[2\mu\omega^2(d_0/3)^3]$. Expanding to
first order in $\epsilon$, we find $\delta x\approx\epsilon
d_0/2$. Taking into account that, if the traps are sufficiently
far apart, the relevant coordinate range is $x_2>x_1$, one
obtains the multipole expansion
\begin{eqnarray}
\frac \lambda {\left| {\mathbf r+ \mathbf d}\right| }&\approx &
\frac \lambda{ d}\left[ 1-\frac x{ d}+\frac{x^2}{ d^2}-\frac
12\frac{y^2}{ d^2}-
\frac 12\frac{z^2}{ d^2}\right.\nonumber\\
&&\qquad\mbox{}+\sum_{k=3}^\infty\left. \frac{P_k(x,y,z)}{
d^k}\right], \label{CoulSeries}
\end{eqnarray}
where each of the multipole terms $P_k$ is a polynomial of
$k^{\rm th}$ degree in $x$, $y$ and $z$ -- for instance,
\begin{subequations}
\begin{eqnarray}
\label{defP3}
P_3(x,y,z)&=&-x\left[x^2-\frac 32(y^2+z^2)\right],\\
P_4(x,y,z)&=&x^4-3x^2(y^2+z^2)+\frac 38
(y^2+z^2)^2.\qquad\label{defP4}
\end{eqnarray}
By virtue of Eq.~(\ref{CoulMin}), the linear term in the expansion
Eq.~(\ref{CoulSeries}) cancels exactly with the one arising from
the harmonic potential $\mu\omega^2(\mathbf r+\mathbf d-\mathbf
d_0)^2/2$ in Eq.~(\ref{defHrApp}). We define the unperturbed
Hamiltonians
\end{subequations}
\begin{align}
H_{R}^0 &\equiv\frac{{\mathbf P}^2}{2M} +\frac 12M\omega
^2{\mathbf R}^2, &H_{r}^0 &\equiv H_x+H_\perp,
\end{align}
where
\begin{subequations}
\begin{eqnarray}
H_x&\equiv &\frac{p_x^2}{2\mu }+\frac 12\mu \nu ^2x^2, \\
H_\perp&\equiv & \frac{p_y^2+p_z^2}{2\mu }+\frac 12\mu \nu _\perp
^2(y^2+z^2);
\end{eqnarray}
\end{subequations}
the higher-multipoles contribution
\begin{equation}
H_1 \equiv \frac \lambda d\sum_{k=3}^\infty
\frac{P_k(x,y,z)}{d^k}; \label{defH1}
\end{equation}
the force terms
\begin{align}
F(t)&=\frac{\hbar\omega}{a_\omega}\bigl(
\hat{\Pi}_1+\hat{\Pi}_2\bigr)\,{\cal F}(t),
&f(t)&=\frac{\hbar\omega}{a_\omega}\frac
{\hat{\Pi}_2-\hat{\Pi}_1}2 \,{\cal F}(t),
\end{align}
where $\hat{\Pi}_i^{\alpha}$ is the projector onto the internal
state $\alpha$ of particle $i$:
\begin{equation}
\hat{\Pi}_1\equiv \left| 1\right\rangle _1\left\langle 1\right|
\otimes {\mathbf 1}_2,\qquad \hat{\Pi}_2\equiv {\mathbf
1}_1\otimes \left| 1\right\rangle _2\left\langle 1\right| ;
\end{equation}
the rescaled frequencies
\begin{align}
\nu & \equiv \omega\sqrt{1+\epsilon}, &\nu _\perp &\equiv
\omega\sqrt{1-\epsilon/2}\;;
\end{align}
and the shifted coordinate and energy scales
\begin{align}
X_0&\equiv {\mathbf R}_0\cdot (1,0,0), & E_0 &\equiv \frac
\lambda d+\frac 12\mu\omega^2\delta x^2.\label{defFf}
\end{align}
Shifting the coordinate system by ${\mathbf R}_0-{\mathbf d}$ and
the energy
scale by $E_0$, we finally obtain Eqs.~(\ref{defHR}) and (\ref{defHr}).

\subsection{Fidelity}
\label{appendix:fidclass}

The goal of this Section is to show the derivation of the
analytical temperature dependence of the fidelity,
Eq.~(\ref{evalFcl4}). We begin by writing the two-particle phase
as
\begin{equation}
\varphi^{\alpha\beta}_{\rm
cl}-\langle\!\langle\varphi^{\alpha\beta}_{\rm
cl}\rangle\!\rangle=\delta_{\alpha\beta}
\left[(-1)^\beta\kappa+1\right]\varepsilon+o[(a_{\tilde\omega}/d)^3],
\end{equation}
where
\begin{subequations}
\begin{eqnarray}
\kappa&\equiv&\frac d{\sqrt 2\xi a_{\tilde\omega}},\\
\varepsilon&=&\frac{3\theta_{\rm cl}}{\hbar\tilde\omega}
\left(\frac{a_{\tilde\omega}}d\right)^2
\Big\{E_1+E_2-2\sqrt{E_1E_2}\cos[\tilde\omega(t_1-t_2)]\nonumber\\
&&\qquad\qquad\qquad\mbox{}-2k_BT\Big\}.
\end{eqnarray}
\end{subequations}
We define
\begin{eqnarray}
\Xi(a,b)&\equiv&\left|\langle\chi_{\rm cl}'|\chi_{\rm
cl}(t)\rangle\right|^2\nonumber\\
&=&2\Big(a\{1-\cos[(\kappa-1)\varepsilon]\}
-b\cos[(\kappa-1)\varepsilon]\Big)\qquad\\
&&\mbox{}\times(a+b-1)+(b-1)^2+
b\left[b+2a\cos(2\kappa\varepsilon)\right],\nonumber
\end{eqnarray}
where $a\equiv|c_{01}|^2$, $b\equiv|c_{10}|^2$, and the
normalization of $|\chi_{\rm cl}\rangle$, in the form
$1=\sum_{\alpha,\beta}|c_{\alpha\beta}|^2$, has been taken into
account. From Eq.~(\ref{defFcl}) it follows
\begin{equation}
F_{\rm cl}=\min_{\{a,b\}}\langle\!\langle\Xi(a,b)
\rangle\!\rangle=
\Big\langle\!\!\Big\langle\min_{\{a,b\}}\Xi(a,b)\Big\rangle\!\!\Big\rangle
\end{equation}
which is a constrained minimization problem, with constraints
$0\leq a\leq 1$, $0\leq b\leq 1$. The solution cannot be found by simply
equating the partial derivatives of $\Xi(a,b)$ to zero, since --~as it will be
seen at the end of this Section~--
the minimum turns out to be located at the border of
the region of allowed parameters. Therefore we must take a closer look at the
problem to find the analytical solution. To this end, we evaluate
\begin{subequations}
\begin{eqnarray}
\partial_a
\Xi(a,b)&=&\!-2\varepsilon^2(\kappa-1)^2
\Big(b\frac{\kappa+1}{\kappa-1}+\frac
12-a\Big)+o[\varepsilon^4],\nonumber\\&&\label{daXi}\\
\partial_b
\Xi(a,b)&=&\!-2\varepsilon^2(\kappa+1)^2\Big(a\frac{\kappa-1}{\kappa+1}+\frac
12-b\Big)+o[\varepsilon^4].\nonumber\\\label{dbXi}
\end{eqnarray}
\end{subequations}
While looking for the minimum, we will neglect
$o[\varepsilon^4]\propto(a_\omega/d)^8$ -- then, to evaluate it, we will use the
exact form of $\Xi(a,b)$.
According to Eq.~(\ref{daXi}), in the region of the parameter plane defined by
the condition
\begin{equation}
\label{bcond}
b\leq \left(a-\frac 12\right)\frac{\kappa-1}{\kappa+1},
\end{equation}
it is $\partial_b\Xi(a,b)\leq-2\kappa(\kappa+1)<0$.
Since $\kappa>1$, the inequality Eq.~(\ref{bcond}) also implies $b<1/2$.
Therefore the minimum must be found outside the region defined by
Eq.~(\ref{bcond}), i.e. for
\begin{equation}
a<b\frac{\kappa+1}{\kappa-1}+\frac 12.
\end{equation}
The latter, by Eq.~(\ref{dbXi}), implies
$\partial_a\Xi(a,b)<0$.
Summing up, the minimum is reached for the values $(a_0,b_0)$ of the parameters,
where
\begin{equation}
a_0=1,\qquad \frac 12\,\frac{\kappa-1}{\kappa+1}<b_0\leq 1.
\end{equation}
The problem is therefore reduced to a one-dimensional constrained minimization:
We have to study the equation
\begin{eqnarray}
0=\partial_b\Xi(a_0,b)&=&2\big(\cos(2\kappa\varepsilon)-
\cos[(\kappa-1)\varepsilon]\nonumber\\
&&\mbox{}-2b\{\cos[(\kappa+1)\varepsilon]-1\}\big),
\qquad\quad
\end{eqnarray}
which has solution
\begin{equation}
\bar b\equiv\frac 12\frac{\sin[(3\kappa-1)\varepsilon/2]}
{\sin[(\kappa+1)\varepsilon/2]}.
\end{equation}
Since the constraint $b_0\leq 1$ has to be fulfilled, we obtain $b_0=\min\{\bar
b,1\}$. Indeed, it is $\bar b\leq 1$ only for $\varepsilon\geq
f_\kappa(\varepsilon)$, where
\begin{equation}
f_\kappa(\varepsilon)\equiv 2\arcsin\left[\frac{\sin(3\kappa\varepsilon/2)-2\sin(\kappa\varepsilon)}
{\sqrt{5+4\cos(2\kappa\varepsilon)}}\right].
\end{equation}
Hence, to $o[(a_\omega/d)^4]$,
\begin{equation}
\label{analFcl}
F_{\rm cl}\approx\left\{\begin{array}{ll}
\displaystyle{\frac 12}\big\{1+
\langle\!\langle\cos[(3\kappa-1)\varepsilon]
\rangle\!\rangle\big\}
\quad&\varepsilon\geq f_\kappa(\varepsilon),\\
1-4\langle\!\langle\cos(\kappa\varepsilon)[\cos(\varepsilon)-
\cos(\kappa\varepsilon)]\rangle\!\rangle&{\rm otherwise.}
\end{array}
\right.
\end{equation}
Now, we $f_\kappa'(0)=\kappa/3>1$ (the inequality following from
$a_\omega\ll d$). Hence for small $T$ --~such that
$\langle\!\langle\varepsilon\rangle\!\rangle\ll 1$, i.e.
$k_BT\ll\hbar\omega(d/a_\omega)^2$~--, it is
$\varepsilon<f_\kappa(\varepsilon)$ and the form of $F_{\rm cl}$
is given by the second row at right-hand side of
Eq.~(\ref{analFcl}), which we can expand in Taylor series around
$T=0$ for taking the thermal average to finally obtain
Eq.~(\ref{evalFidcl}). When $\kappa=0$, i.e. the third-order
anharmonic correction is suppressed,
\begin{equation}
\label{Xi4}
\Xi(a,b)\Big|_{\kappa=0}=1+2(a+b-1)(a+b)[1-\cos(\varepsilon)].
\end{equation}
The function to be minimized depends now only on the sum $a+b$. Therefore the
minimum can be searched by fixing one of the two parameters and varying only the
other one. We can choose $a_0=1$ as before, and minimize Eq.~(\ref{Xi4})
with respect to $b$.
Eq.~(\ref{analFcl}) gives the solution also in this case. In
particular, since $f_{\kappa=0}(\varepsilon)\equiv0<\varepsilon$
$\forall\varepsilon$,
the analytical expression for the fidelity is
now given by the first row at right-hand side of Eq.~(\ref{analFcl}), which for
$\kappa=0$ becomes simply
\begin{equation}
\label{analFcl4}
F_{\rm cl}'\equiv F_{\rm cl}\big|_{\kappa=0}\approx
\frac 12\big[1+
\langle\!\langle\cos(\varepsilon)\rangle\!\rangle\big].
\end{equation}
The same procedure can be used for the minimization over the
possible internal states also in the quantum case, as is done in
App.~\ref{appendix:fid}. By expanding Eq.~(\ref{analFcl4}) to
lowest nonzero order in powers of $a_\omega/d$, we obtain finally
Eq.~(\ref{evalFcl4}).

\section{Quantum calculation}

In this Appendix we compute both analytically and numerically the
evolution of the two-ion system, evaluate the resulting phase
shifts and derive an accurate expression for the fidelity, giving
as well the explicit expression of some quantities used in the
text.

\subsection{Undoing single-particle phases}
\label{appendix:single}

In this Section we show how to get rid of the spurious phases
accumulated during gate operation, in order to be left with the
gate phase Eq.(\ref{defvartheta}). In the ideal case where the
external degrees of freedom factorize out at the end of the
computation, the evolution operator Eq.~(\ref{defUt}) induces both
two-particle conditional phases and single-particle kinetic
phases, depending on each ion's external state. When both
particles are in their external ground state, we can undo the
kinetic phases and other inessential phases through single-bit
operations of the form
\begin{equation}
\label{defSj}
S_j=e^{-i\varphi_j}\sum_{\alpha}|\alpha\rangle_j\langle\alpha|\;e^{is^\alpha_j},
\end{equation}
where $\varphi_j$ has to be equal to the kinetic phase acquired
after the gate operation by particle $j$, and we choose
\begin{equation}
\begin{aligned}
s_1^0&=-\varphi^{00}/2,& \quad s_1^1&=-\varphi^{10}+s_1^0;\\
s_2^0&=s_1^0,& s_2^1&=-\varphi^{01}+s_1^0.
\end{aligned}
\end{equation}
Under these conditions, the compound operator
\begin{equation}
{\cal U}(t)\equiv \left(S_1\otimes S_2\right)U(t,t_0)
\end{equation}
implements the transformation Eq.~(\ref{defGate}), with
$\vartheta$ given by Eq.~(\ref{defvartheta}).

\subsection{Unperturbed solution}
\label{appendix:unpert}

As explained in the text, since we assume $d\gg a_\omega$ we can
treat the higher-multipoles term $H_1$ as a small perturbation
with respect to the rest of the Hamiltonian. In this Section we
solve exactly the unperturbed problem, i.e. we calculate the
time-dependent evolution dictated by $H(t)-H_1$. So we want to
solve the Schr\"{o}dinger equation
\begin{equation}
\label{Schroed} i\hbar |\dot\Psi (t)\rangle
=\left[H_R(t)+H_r^0(t)-{\bf f}(t)\cdot({\bf r}+{\bf
d})\right]|\Psi (t)\rangle,
\end{equation}
with initial condition
\begin{equation}
\left| \Psi(t_0)\right\rangle \equiv \left|\psi_R(t_0)\right\rangle _R
\left| \psi_r(t_0)\right\rangle_r \left|
\alpha \right\rangle _1\left| \beta \right\rangle _2.
\label{Psit0}
\end{equation}
The subscript $R$ ($r$) denotes the center-of-mass (relative)
motion, as defined in App.~\ref{appendix:hamil}. The solution is
\cite{Galindo}
\begin{subequations}
\begin{eqnarray}
\left| \psi _{R}(t)\right\rangle _R &=&e^{-iH_{{R}
}^0(t-t_0)/\hbar }e^{iX_0\int_{t_0}^tF(t^{\prime })dt^{\prime
}/\hbar }
\label{psiCManal}\\
&&\times \exp\left[ -\int_{t_0}^tds\,K_{R}(s,t_0)\frac{dK_{
{R}}^{*}(s,t_0)}{ds}\right]\nonumber\\
&&\times e^{-iK_{ {R}}(t,t_0)\hat a_R^{\dagger }}e^{-iK_{{R}
}^{*}(t,t_0)\hat a_R}  \left| \psi _{{R}
}(t_0)\right\rangle_R ,  \nonumber\\
\left| \psi _{r}(t)\right\rangle_r  &=&e^{-iH_{{r}
}^0(t-t_0)/\hbar }e^{id\int_{t_0}^tf(t^{\prime })dt^{\prime
}/\hbar}
\label{psiRelanal}\\
&&\times \exp\left[ -\int_{t_0}^tds\,K_{r}(s,t_0)\frac{
dK_{r}^{*}(s,t_0)}{ds}\right]\nonumber\\
&&\times e^{-iK_{r}(t,t_0)\hat a_r^{\dagger }}e^{-iK_{{r}
}^{*}(t,t_0)\hat a_r}   \left| \psi _{{r} }(t_0)\right\rangle
_r,  \nonumber
\end{eqnarray}
where $\hat a_R$ ($\hat a_r$) is the annihilation operator for the
$x$ component of the center-of-mass (relative) motion, and
\end{subequations}
\begin{subequations}
\begin{eqnarray}
K_{R}(t,t_0) &\equiv &\frac 1{\sqrt{2M\hbar\omega} }
\int_{t_0}^tdt'\,F(t')e^{i\omega (t'-t_0)},\qquad \label{defKCM}\\
K_{r}(t,t_0) &\equiv &\frac 1{\sqrt{2\mu\hbar\nu}}
\int_{t_0}^tdt'\,f(t')e^{i\nu (t'-t_0)}.\label{defKrel}
\end{eqnarray}
Now the explicit form of the force term, Eq.~(\ref{defcalF}),
can be inserted into Eqs.~(\ref{defKCM}) and (\ref{defKrel}) through
Eq.~(\ref{defFf}), to yield
\end{subequations}
\begin{subequations}
\begin{eqnarray}
K_{R}(t,t_0)&=&\left( \hat{\Pi}_1+\hat{\Pi}_2\right) K\left( \omega
,t,t_0\right) ,\\
K_{r}(t,t_0)&=&\left( \hat{\Pi}_2-\hat{\Pi} _1\right) K\left( \nu
,t,t_0\right) ,
\end{eqnarray}
where
\end{subequations}
\begin{subequations}
\begin{eqnarray}
K\left( \omega ,t,t_0\right)  &\equiv &\frac{ \sqrt{\pi}}4\omega
\tau \xi \left. e^{-\left( \omega \tau /2\right) ^2}I\left( \omega
,t^{\prime }\right) \right| _{t_0}^t,\qquad \\
I\left( \omega ,t\right)  &\equiv &\mathrm{Erf}\left( \frac t\tau
-\frac{ i\omega \tau }2\right) .
\end{eqnarray}
We take as initial state
\end{subequations}
\begin{subequations}
\begin{eqnarray}
\left| \psi _R(t_0)\right\rangle_R&\equiv &
\left| {\bf N}\right\rangle _R  \equiv  \left| N_X\right\rangle
_X\left| N_Y\right\rangle _Y\left| N_Z\right\rangle _Z,\qquad \\
\left| \psi _r(t_0)\right\rangle_r&\equiv &
\left| {\bf n}\right\rangle_r  \equiv
\left| n_x\right\rangle _x\left|
n_y\right\rangle _y\left| n_z\right\rangle _z.
\end{eqnarray}
During time evolution, the two ions
will acquire a state-dependent phase shift
\end{subequations}
\begin{equation}
\label{Over0}
\langle\Psi(t)|\Psi(t_0)\rangle\equiv
\left|\langle\Psi(t)|\Psi(t_0)\rangle\right|e^{i\phi^{\alpha\beta}},
\end{equation}
which turns out to be given by
\begin{subequations}
\begin{eqnarray}
\phi^{\alpha\beta}&=&
\phi^{\alpha\beta}_{R}+\phi^{\alpha\beta}_{r},\label{phiab}\\
\phi^{\alpha\beta} _{R} &\approx &
\left(\alpha+\beta\right)^2\xi ^2
\Phi \left( \omega \right) +\left(\alpha+\beta\right)
\sqrt{\pi }\omega\tau\xi X_0/a_\omega\nonumber\\
&&\quad-(N_X+N_Y+N_Z)\omega \left( t-t_0\right),\label{phiabCM} \\
\phi^{\alpha\beta} _{r} &\approx &
\left(\alpha-\beta\right)^2\xi ^2
\Phi \left( \nu\right)-\left(\alpha-\beta\right)
\sqrt{\pi }\omega\tau\xi d/a_\omega\nonumber\\
&&\quad-\left[n_x\nu+(n_y+n_z)\nu_\perp\right] \left( t-t_0\right),
\label{phiabRel}\\
\Phi( \omega)  &\equiv &-\pi\omega^2\tau
\frac {A^2}{8C}D(t')\Big|_{t_0}^te^{(B/C)^2-(\omega \tau /2) ^2},\label{Phiab}
\end{eqnarray}
\end{subequations}
with
\begin{eqnarray}
A &\equiv &\Im \left[ I\left( \omega ,t_0\right) \right] -iI^{*}\left(
\omega ,0\right) , \nonumber\\
B &\equiv &\omega\Re \left[ I\left( \omega ,t_0\right) \right] , \nonumber\\
C &\equiv &2\sqrt{A^2\left( \frac{\omega ^2}2+\frac 1{\tau ^2}\right) -\frac
{A\omega}{\sqrt{\pi }\tau}e^{\left( \omega \tau /2\right) ^2}+
\frac{B^2}4},\nonumber\\
D(t)&\equiv &\mathrm{Erf}\left( \frac BC+\frac{C}{2A}t\right).
\end{eqnarray}
The equality in Eqs.~(\ref{phiabCM}) and (\ref{phiabRel}) is approximate since
the integrals in the exponent of Eqs. (\ref{psiCManal}) and (\ref
{psiRelanal}) have been evaluated by means of a saddle-point approximation,
giving a very good agreement (relative difference less than $10^{-5}$ with
typical parameters as used here) with the exact result, which cannot be
evaluated analytically.
Finally, from Eqs. (\ref{psiCManal}) and (\ref
{psiRelanal}) we obtain
\begin{equation}
{\cal O}^{\{n_i\}}(t,t_0)\equiv\left|\langle\Psi(t)|\Psi(t_0)\rangle \right| =
{\mathcal O}^{\{n_i\}}_{R}\,{\mathcal O}^{\{n_i\}}_{r}, \label{OROr}
\end{equation}
where
\begin{subequations}
\begin{eqnarray}
{\mathcal O}^{\{n_i\}}_{R}&=&M\bigl( -N_X,1,\left| K_{{R}
}(t,t_0)\right| ^2\bigr) e^{-\frac 12 \left| K_{R}(t,t_0)\right|
^2},\qquad\quad\label{defOR}\\
{\mathcal O}^{\{n_i\}}_{r}&=&M\bigl( -n_x,1,\left| K_{{r}
}(t,t_0)\right| ^2\bigr) e^{-\frac 12\left| K_{r}(t,t_0)\right|
^2},\qquad\quad\label{defOr}
\end{eqnarray}
and $M(a,b,z)$ is the \textit{confluent hypergeometric function}.
\end{subequations}

\subsection{First-order perturbation theory}
\label{appendix:pert}

Now we want to evaluate the lowest-order corrections that appear
when the higher multipole contributions in the Hamiltonian are
taken into account. Following \cite{Galindo}, we expand the
evolution operator as
\begin{equation}
U\left( t,t_{0}\right) =U_{0}\left( t,t_{0}\right) +\sum_{j=1}^{\infty
}U_{j}\left( t,t_{0}\right) ,  \label{UPert}
\end{equation}
where $U_{0}\left( t,t_{0}\right) $ is the operator of the unperturbed
evolution, already calculated in App.~\ref{appendix:unpert}, and
\begin{eqnarray}
U_{j}\left( t,t_{0}\right)  &\equiv &\frac{1}{(i\hbar )^{j}}
\int_{t_{0}}^{t}dt_{j}\int_{t_{0}}^{t_{j}}dt_{j-1}\ldots
\int_{t_{0}}^{t_{2}}dt_{1}U_{0}\left( t,t_{j}\right)   \nonumber \\
&&\times H_{1}U_{0}\left( t_{j},t_{j-1}\right) H_{1}U_{0}\left(
t_{j-1},t_{j-2}\right)  \\
&&\times\ldots U_{0}(t_{2},t_{1})H_{1}U_{0}(t_{1},t_{0}).  \nonumber
\end{eqnarray}
We are interested in evaluating the diagonal matrix elements
$\left\langle \Psi
(t_{0})\right| U(t,t_{0})\left| \Psi(t_{0})\right\rangle $ to first order,
according to Eq. (\ref{UPert}). Since $\left\langle
U_{0}(t,t_{0})\right\rangle $ is given by Eq.~(\ref{Over0}), we just
need to compute
\begin{eqnarray}
\langle U_{1}(t,t_0)\rangle
&=&\frac{1}{i\hbar }\int_{t_{0}}^{t}dt^{\prime }{\cal O}_{1}^{\alpha \beta
}(t,t^{\prime },t_{0})e^{i[\phi _{r}^{\alpha \beta }(t,t^{\prime })+\phi
_{r}^{\alpha \beta }(t^{\prime },t_{0})]}\nonumber \\
&=&\frac{e^{i\phi _{r}^{\alpha \beta }(t,t_{0})}}{i\hbar }
\int_{t_{0}}^{t}dt^{\prime }{\cal O}_{1}^{\alpha \beta
}(t,t^{\prime },t_{0})
\end{eqnarray}
where the unperturbed phase factorizes since (as shown in
Sec.~\ref{section:quadr}) it does not depend on the initial state, and we have
defined
\begin{equation}
{\cal O}_{1}^{\alpha \beta }(t,t^{\prime },t_{0})\equiv \left| \left\langle
\Psi(t_{0})\right| U_{0}(t,t^{\prime })H_{1}U_{0}(t^{\prime },t_{0})\left|
\Psi(t_{0})\right\rangle \right| .
\end{equation}
The exact result, given by Eqs.~(\ref{psiCManal}) and
(\ref{psiRelanal}), cannot be integrated analytically over time.
Instead we adopt the adiabatic approximation, i.e. we assume that
the condition (i) of Sect.~\ref{section:class} is satisfied. The
Hamiltonian then changes slowly enough so that the system, being
in a motional eigenstate at $t=t_0$, follows the changes being in
the corresponding eigenstate at every subsequent time $t$. This
means in our case that, if $t_0<0$, $t>0$ and their absolute
values are large enough, we will have
$|\Psi(t)\rangle\approx|\Psi(t_0)\rangle$. The relative-motion
wavefunction of the evolved state is then
\begin{eqnarray}
|\langle \psi_r(t_{0})| U_{0}(t,t^{\prime })|{\bf r}\rangle|
&\approx &|\langle {\bf r}| U_{0}(t^{\prime },t_{0})
| \psi_r(t_{0})\rangle |\\
&\approx&\psi_{n_x}\big(x-f(t)/\mu\nu^2\big)
\psi_{n_y}(y)\psi_{n_z}(z),
\nonumber
\end{eqnarray}
where e.g. $\psi_{n_x}(x)\equiv\langle x|n_x\rangle_x\in {\rm l\!R}$.
Finally we obtain
\begin{equation}
\label{defU1}
\langle U_{1}(t,t_0)\rangle\approx i e^{i\phi _{r}^{\alpha \beta }}
\left(\Delta^{\alpha\beta}+\Delta'\right),
\end{equation}
where
\begin{subequations}
\begin{eqnarray}
\Delta^{\alpha\beta}&\equiv &-\frac{\sqrt{\pi}\tau}\hbar\frac\lambda d
\sum_{k=3}^\infty\left[\frac{a_\nu}d\tilde\xi(\alpha-\beta)\right]^k\delta_k,
\label{defDeltaab}\\
\Delta'&\equiv &-\frac {t-t_0}\hbar\frac\lambda
d\sum_{k=3}^\infty\left(\frac{a_\nu}{d}\right)^k\delta_k',\\
\delta_k&\equiv &\frac 1{\sqrt{\pi}\tau(a_\nu\tilde\xi)^k}\int_{t_0}^tdt'
\langle{\bf n}|\Big[P_k\Big(x+\frac{\omega^2}{\nu^2}a_\omega
{\cal F}(t),y,z\Big)\nonumber\\
&&\qquad\qquad\qquad\qquad\quad\mbox{}-P_k(x,y,z)\Big]|{\bf n}\rangle_r,\\
\delta_k'&\equiv &\frac {\langle{\bf n}|
P_k(x,y,z)|{\bf n}\rangle_r}{a_\nu^k},\label{defd1}\\
\tilde\xi&\equiv &\xi
\frac{a_\nu}{a_\omega}\frac\omega\nu=\frac{\sqrt{2}\xi}{(1-\epsilon)^{3/4}},
\qquad a_\nu \equiv \sqrt{\hbar /\mu \nu }.
\end{eqnarray}
From Eqs.~(\ref{UPert}) and (\ref{defU1}) it follows that, to first order,
\end{subequations}
\begin{eqnarray}
\langle U(t,t_0)\rangle&\approx &
\langle U_0(t,t_0)+ U_1(t,t_0)\rangle\nonumber\\
&=&\langle U_0(t,t_0)\rangle\left[1+
i\frac{\Delta^{\alpha\beta}+\Delta'}
{|\langle U_0(t,t_0)\rangle|}\right],\label{phipert1}
\end{eqnarray}
which is equivalent to Eq.~(\ref{Upert1}), given that
$|\Delta^{\alpha\beta}+\Delta'|= |\langle
U_1(t,t_0)\rangle|\ll|\langle U_0(t,t_0)\rangle|\approx 1$. The
internal-state-independent part $\Delta'$ cancels out when
computing the gate phase Eq.~(\ref{defvartheta}), as well as the
terms of odd $k$ in $\Delta^{\alpha\beta}$, due to the summation
over the internal states. The adimensional quantities $\delta_k$
and $\delta_k'$ do not depend either on the internal state nor on
time, but just on the relative motional state. We will now
calculate them for $k=3,4$. To be precise, we should not use the
eigenstates $|{\bf n}\rangle_r$ of $H_r^0$, as is done in
Eq.~(\ref{defd1}), but rather those of the full Hamiltonian
$H_r$. However, as we will demonstrate in the next Section, the
corrections are of $o[(a_\omega/d)^3]$ and therefore we will
consistently not take them into account in the present
calculation. The relevant matrix elements are
\begin{subequations}
\begin{eqnarray}
\left\langle n\right| x\left| n^{\prime }\right\rangle  &=&
\frac{a_\nu }{\sqrt{2}}\left( \delta _{n^{\prime
},n-1}\sqrt{n}+\delta _{n^{\prime },n+1}\sqrt{n+1}\right) , \\
\left\langle n\right| x ^2\left| n^{\prime }\right\rangle
&=&\frac{ a_\nu ^2}2\left[ \delta _{n^{\prime },n-2}\sqrt{n\left(
n-1\right) }+\delta
_{n^{\prime },n}\left( 2n+1\right) \right.\nonumber\\
&&\left.\mbox{}+\delta _{n^{\prime },n+2}\sqrt{\left(
n+1\right) \left( n+2\right) }\right] , \\
\left\langle n\right| x^3\left| n^{\prime }\right\rangle
&=&\frac{a_\nu ^3}{ 2^{3/2}}\left[ \delta _{n^{\prime
},n-3}\sqrt{n\left( n-1\right) \left(
n-2\right) }\right.\nonumber\\
&&\mbox{}+3\delta _{n^{\prime },n-1}n
+3\delta _{n^{\prime },n+1}\left( n+1\right) ^{3/2} \\
&&\left. +\delta
_{n^{\prime },n+3}\sqrt{\left( n+1\right) \left( n+2\right) \left(
n+3\right) }\right] \nonumber,\\
\left\langle n\right| x^4\left| n^{\prime }\right\rangle
&=&\frac{a_\nu ^4}4 \left\{ \delta _{n^{\prime
},n-4}\sqrt{n\left( n-1\right) \left( n-2\right)
(n-3)}\right.  \nonumber\\
&&+2\delta _{n^{\prime },n-2}(2n+1)\sqrt{n\left( n-1\right) }\nonumber\\
&&+3\delta _{n^{\prime },n}[2n(n+1)+1]\\
&&+2\delta _{n^{\prime },n+2}(2n+3)\sqrt{
\left( n+1\right) (n+2)} \nonumber\\
&&+\left. \delta _{n^{\prime },n+4}\sqrt{\left( n+1\right) \left( n+2\right)
\left(n+3\right) (n+4)}\right\} \!,\nonumber
\end{eqnarray}
Hence
\end{subequations}
\begin{subequations}
\begin{eqnarray}
\delta_3&= &-\frac{1}{\sqrt{3}}-\frac{3}{2\tilde\xi^2}
\left[2n_x+1-\tilde\nu(n_y+n_z+1)\right],\\
\delta_4&= &\frac 12+\frac{3}{\sqrt{2}\tilde\xi^2}
\left[2n_x+1-\tilde\nu(n_y+n_z+1)\right],\label{eta4}\\
\delta_3'&=&0,\\
\delta_4'&= &\frac 34
\left[2n_x(n_x+1)+1\right]-\frac{3}2\tilde\nu(2n_x+1)(n_y+n_z+1)\nonumber\\
&&\mbox{}+\frac{3}{16}\tilde\nu^2
\left[n_y(3n_y+5)+n_z(n_z+5)+4(1+n_yn_z)\right]\nonumber\\\label{defd41}
\end{eqnarray}
where $\tilde\nu\equiv\nu/\nu_\perp$.
\end{subequations}

\subsection{Perturbative corrections to the eigenstates}
\label{appendix:statepert}

Since in our case the perturbation $H_1$ is static, its effect on
the initial eigenstates of the system must be taken into account.
In this Section we show how to do that in second-order
perturbation theory. Our problem is to compute the eigenstates of
the initial relative-motion Hamiltonian
\begin{equation}
H_r(t_0)=H_r^0+H_1=H_r^0+\epsilon H_r^1,
\end{equation}
whereby the external force is vanishing at the initial time, and
\begin{equation}
H_r^1\equiv\frac{\hbar\omega}2\frac{d^2}{a_\omega^2}
\sum_{k=3}^\infty\frac{P_k(x,y,z)}{d^k}.
\end{equation}
Therefore we make a perturbative expansion in the small parameter
$\epsilon$. So we write the eigenstates of $H_r$ (omitting
throughout this Section the subscript $r$ as
\begin{equation}
|{\bf n}(\epsilon)\rangle=\sum_{i=0}^\infty\epsilon^i|{\bf
n}^{(i)}\rangle,
\end{equation}
where the first terms are
\begin{subequations}
\begin{eqnarray}
|{\bf n}^{(1)}\rangle&=&\sum_{{\bf m}\not={\bf n}}
\frac{\langle{\bf m}^{(0)}|H^1_r|{\bf n}^{(0)}\rangle}{E_{\bf
n}^{(0)}-E_{\bf m}^{(0)}} |{\bf m}^{(0)}\rangle,\\
|{\bf n}^{(2)}\rangle&=&\sum_{{\bf l},{\bf m}\not={\bf n}}
\frac{\langle{\bf m}^{(0)}|H^1_r|{\bf l}^{(0)}\rangle \langle{\bf
l}^{(0)}|H^1_r|{\bf n}^{(0)}\rangle}{\big(E_{\bf n}^{(0)}-E_{\bf
m}^{(0)}\big)\big(E_{\bf n}^{(0)}-E_{\bf l}^{(0)}\big)} |{\bf
m}^{(0)}\rangle,\nonumber\\
\end{eqnarray}
\end{subequations}
and $|{\bf n}^{(0)}\rangle$ are the eigenstates of $H_r^0$, with
eigenenergies $E_{\bf n}^{(0)}$. The $k^{\rm th}$ term in $H_r^1$
gives a contribution of order
$\sim(\hbar\omega/2)(a_\omega/d)^{k-2}$. Since we want to neglect
corrections of order $o[(a_\omega/d)^3]$, we need to go up to
$k=4$ in the expansion of $H_r^1$. But from Eqs.~(\ref{defP3})
and (\ref{defP4}) it is straightforward to see that $\langle{\bf
m}^{(0)}|P_{3,4}(x,y,z)|{\bf n}^{(0)}\rangle$ for ${\bf
m}\not={\bf n}$. It follows that
\begin{equation}
|{\bf n}(\epsilon)\rangle=|{\bf n}^{(0)}\rangle+o[(a_\omega/d)^3],
\end{equation}
and therefore, as already anticipated in the previous Section,
for the purpose of the present calculation we can consistently use
the eigenstates of the unperturbed Hamiltonian $H_r^0$.

\subsection{Numerical computation}
\label{appendix:num}

The goal of this Section is to transform the Schr\"odinger
equation for the two-particle wavefunction into a system of
first-order differential equations for the time dependence of its
projections over the initial eigenstates, better suitable for
numerical handling. Since the problem has cylindrical symmetry
around the $x$ axis, the transverse coordinates always appear as
powers of $\rho \equiv \sqrt{y^2+z^2}$. Thus the original
three-dimensional problem is equivalent to a two-dimensional one.
We expand the wavefunction (omitting for simplicity the subscript
$r$) as
\begin{eqnarray}
|\psi (t)\rangle &=&\sum_{n,l=0}^\infty c_{nl}(t)\exp\left\{
\frac i\hbar \left[d\int_{t_0}^tf(t') dt^{\prime }\right.\right.\nonumber\\
&&\qquad\qquad\quad\mbox{}-\hbar( n\nu +l\nu _\perp +1)
t\bigg]\bigg\} \left|nl\right\rangle ,\qquad
\end{eqnarray}
where $\left| nl\right\rangle \equiv \left| n\right\rangle _x\left|
l\right\rangle _\perp $, the $\left| n\right\rangle _x$ ($\left|
l\right\rangle _\perp $) are the eigenstates of $H_x$ ($H_\perp $).
From Eq. (\ref{Schroed}) it follows
\begin{eqnarray}
\dot{c}_{nl}&=&\frac i\hbar \sum_{n^{\prime },l^{\prime }=0}^\infty
c_{n^{\prime }l^{\prime }}(t)e^{i\left[(n-n')\nu+(l-l')\nu_\perp \right]t }
\nonumber\\
&&\quad\times\left\langle nl\right| \left[ f(t)x-H_1\right]
\left| n^{\prime }l^{\prime }\right\rangle\nonumber\\
&=&\frac i\hbar \bigg[\frac{a_\nu }{\sqrt{2}}f(t)\left(\sqrt{n}
e^{i\nu t}c_{n-1,l}+\sqrt{n+1}e^{-i\nu t}c_{n+1,l}\right) \nonumber \\
&&\qquad\mbox{}+\frac\lambda d\sum_{k=3}^\infty
\left(\frac{a_\nu}{\sqrt 2d}\right)^kC_{nl}^{(k)}\bigg], \label{RedRel}
\end{eqnarray}
where the coefficients $C_{nl}^{(k)}$ correspond to the
$k^{\rm th}$ term in Eq.~(\ref{defH1}) -- in particular,
\begin{subequations}
\begin{eqnarray}
C_{nl}^{(3)}&=&\sqrt{n(n-1)(n-2)}e^{i3\nu t}c_{n-3,l} \\
&&+3n^{3/2}e^{i\nu t}c_{n-1,l}+3(n+1)^{3/2}e^{-i\nu t}c_{n+1,l}  \nonumber \\
&&+\sqrt{(n+1)(n+2)(n+3)}e^{-i3\nu t}c_{n+3,l}  \nonumber \\
&&-\frac {3\tilde\nu}2\left[ \sqrt{nl(l-1)}e^{i(\nu +2\nu
_\perp )t}c_{n-1,l-2}\right.   \nonumber\\
&&+\sqrt{n(l+1)(l+2)}e^{i(\nu -2\nu _\perp )t}c_{n-1,l+2}  \nonumber \\
&&+(2l+1)\left( \sqrt{n}e^{i\nu t}c_{n-1,l}+\sqrt{n+1}e^{-i\nu
t}c_{n+1,l}\right)   \nonumber \\
&&+\sqrt{(n+1)l(l-1)}e^{-i(\nu -2\nu _\perp )t}c_{n+1,l-2}  \nonumber\\
&&\left. +\sqrt{(n+1)(l+1)(l+2)}e^{-i(\nu +2\nu _\perp
)t}c_{n+1,l+2}\right];  \nonumber
\end{eqnarray}
\begin{eqnarray}
C_{nl}^{(4)}&=&-\sqrt{n(n-1)(n-2)(n-3)}e^{4i\nu t}c_{n-4,l}\\
&&+\sqrt{n(n-1)}e^{2i\nu t}\left\{3\tilde\nu \left[
\sqrt{l(l-1)}e^{2i\nu_{\perp }t}c_{n-2,l-2} \right.\right.\nonumber\\
&& +\sqrt{(l+1)(l+2)}e^{-2i\nu_{\perp }t}c_{n-2,l+2}\nonumber\\
&&+(2l+1)c_{n-2,l}\Big]
-2(2n-1)c_{n-2,l}\Big\} \nonumber\\
&&-\frac{3\tilde\nu^2}{8}\sqrt{l(l-1)(l-2)(l-3)}
e^{4i\nu _{\perp }t}c_{n,l-4} \nonumber\\
&&+3\tilde\nu\sqrt{l(l-1)}\left[ (2n+1)-\frac{\tilde\nu}4
(2l-1)\right] e^{2i\nu _{\perp }t}c_{n,l-2}\nonumber\\
&&-\left\{ \frac{9\tilde\nu^2}{8}\left[ 2l(l+1)+1
\right] -3\tilde\nu(2n+1)(2l+1)\right.\nonumber\\
&&+3\left[ 2n(n+1)+1\right]
\bigg\} c_{nl} -\frac 34\tilde\nu\sqrt{(l+1)(l+2)}\nonumber\\
&&\qquad\times[\tilde\nu
(2l+3)-4(2n+1)] e^{-2i\nu _{\perp }t}c_{n,l+2} \nonumber\\
&&-\frac{3\tilde\nu^2}{8}\sqrt{(l+1)(l+2)(l+3)(l+4)}
e^{-4i\nu _{\perp }t}c_{n,l+4} \nonumber\\
&&+\sqrt{(n+1)(n+2)}e^{-2i\nu t}\Big\{3\tilde\nu\Big[(2l+1)c_{n+2,l}\nonumber\\
&&+\sqrt{(l+1)(l+2)}e^{-2i\nu _{\perp }t}c_{n+2,l+2}\nonumber\\
&&+\sqrt{l(l-1)}e^{2i\nu_{\perp }t}c_{n+2,l-2}\Big]-2(2n+3)c_{n+2,l}\Big\}
\nonumber\\
&&-\sqrt{(n+1)(n+2)(n+3)(n+4)}e^{-4i\nu t}c_{n+4,l}. \nonumber
\end{eqnarray}
Excitations higher than a certain level should be absent as long
as we are in an adiabatic regime. Thus in Eq.~(\ref {RedRel}) we
neglect the coefficients above a certain $N$. We have checked
that the result is independent of the cutoff.
\end{subequations}

\subsection{Fidelity}
\label{appendix:fid}

The goal of this Section is to evaluate the gate operation
fidelity in the full three-dimensional quantum-mechanical
framework. The overall phase $\Theta$ appearing in
Eq.~(\ref{defgateG}) can be computed from
Eqs.~(\ref{phiab})-(\ref{Phiab}) and
(\ref{defDeltaab})-(\ref{defd1}), as
\begin{equation}
\Theta\approx2\omega\Bigg\{\sqrt\pi\xi\tau\frac{X_0}{a_\omega}-\bigg[
\sum_{i=1}^6n_i+\epsilon\sum_{k=3}^\infty
\Big(\frac{a_\omega}d\Big)^k\delta_k'\bigg](t-t_0)\!\Bigg\}\!,
\end{equation}
where $\delta_4'$ is defined in Eq.~(\ref{defd41}), and it has
been taken into account that $\epsilon\ll 1$. In the ideal case,
according to Eq.~(\ref{defGate}) for $\vartheta=\pi$, the gate
operation transforms the initial internal state $|\chi\rangle$
into
\begin{equation}
|\chi'\rangle=\sum_{\alpha,\beta=0}^1(-1)^{\alpha\beta}
c_{\alpha\beta}|\alpha\rangle_1|\beta\rangle_2.
\end{equation}
In a more realistic situation the initial total density operator $\sigma_T$ at a
temperature $T$ is given by
\begin{equation}
\sigma_T=\rho_T(t_0)\otimes|\chi\rangle\langle\chi|,
\end{equation}
where $\rho_T(t_0)$ is defined in Eq.~(\ref{defrhoT}), and we recall that
$\omega\approx\nu\approx\nu_\perp$. After the gate operation we have
\begin{equation}
\sigma_T'=\!\!\sum_{\alpha,\beta,\alpha',\beta'}
c_{\alpha\beta}c^*_{\alpha'\beta'}
G_{\alpha\beta}\,\rho_T(t_0)G_{\alpha'\beta'}^\dagger
|\alpha\rangle_1\langle\alpha'|\otimes|\beta\rangle_2\langle\beta'|,
\end{equation}
where $G_{\alpha\beta}\equiv\langle\alpha\beta|G|\alpha\beta\rangle$, and the
gate operator $G$ is defined in Eq.~(\ref{defgateG}).
As already stated in Sect.~\ref{section:quadr}, because of adiabaticity,
the motional state after the gate operation is unchanged, i.e.
$G_{\alpha\beta}\,\rho_T(t_0)G_{\alpha'\beta'}^\dagger\approx\rho_T(t_0)$.
If $\theta=\pi$, the minimum fidelity $F(T)$, given by Eq.~(\ref{defFid}), is
\begin{eqnarray}
F(T)&=&\min_{\{c_{\alpha\beta}\}}(1-\gamma)^6\prod_{i=1}^6\sum_{n_i=1}^\infty\gamma^{n_i}
\langle n_i|\Big[\big(|c_{00}|^2+|c_{11}|^2\big)^2\nonumber\\
&&\mbox{}+2\big(|c_{00}|^2+|c_{11}|^2\big)\big(|c_{01}|^2+|c_{10}|^2\big)
\cos(\Delta\theta)\nonumber\\
&&\hspace{3truecm}\mbox{}+\big(|c_{01}|^2+|c_{10}|^2\big)^2\Big]|n_i\rangle\nonumber\\
&=&\frac{(1-\gamma)^3}2\prod_{i=1}^3\sum_{n_i=1}^\infty\gamma^{n_i}\langle n_i|
[1+\cos(\Delta\theta)]|n_i\rangle\\
&\approx&1-\frac{6^3\theta_{\rm
cl}^2}{(1+\epsilon^5)(1-\epsilon^2/4)}
\left(\frac{a_\omega}{d}\right)^4\frac{e^{-\hbar\nu/k_BT}}
{(1-e^{-\hbar\nu/k_BT})^2}\nonumber \label{minFidQ}
\end{eqnarray}
where the minimization over the coefficients $\{c_{\alpha\beta}\}$
has been carried out exactly as in
App.~\ref{appendix:fidclass}. Here,
only the relative motion comes into play because
$\Delta\theta$ is independent of the center-of-mass motion, and
$\cos(\Delta\theta)$ has been expanded up to $o(\Delta\theta^3)$.
Hence Eq.~(\ref{evalF}) follows, by taking into account that
$\epsilon\ll 1$ and $\theta_{\rm
cl}\approx\theta=\pi$, and expanding in a Taylor series for
$\hbar\omega\ll k_BT$.

\centerline{}
\subsection{Many-ions calculation}
\label{appendix:many}

In this Section we simply give the definitions of the parameters
appearing in Eq.~(\ref{HNions}):
\begin{subequations}
\begin{eqnarray}
\varepsilon_i&\equiv &
\frac{\omega_i^2}{\omega^2}\,\tilde x_i^2-\frac\epsilon 2\,d^2{\cal H}_{n-i},\\
\omega_i&\equiv &\omega\sqrt{1+\epsilon\eta_i},\qquad
\tilde x_i\equiv\frac d2\frac{\epsilon\eta_i'}{1+\epsilon\eta_i},\\
\eta_i&\equiv &\frac 12\sum_{j=1}^N\frac {1-\delta_{ij}}{|i-j|^3}\nonumber\\
&=&\frac 14\left[\psi^{(2)}(i)+\psi^{(2)}(N+1-i)+\zeta(3)\right],\\
\eta_i'&\equiv &\frac 12\sum_{j=1}^N\frac {i-j}{|i-j|^3}\nonumber\\
&=&\frac 12\left[\psi^{(1)}(N+1-i)-\psi^{(1)}(i)\right],
\end{eqnarray}
where ${\cal H}_k$ is the {\em harmonic number} and $\psi^{(k)}(z)$ the {\em
polygamma function} of order $k$, and $\zeta(s)$ is the {\em Riemann zeta
function}. It is
\end{subequations}
\begin{equation}
\max_{i,n}|\eta_i|=\zeta(3)\approx 1.2,\qquad
\max_{i,n}|\eta_i'|=\frac{\pi^2}{12}\approx 0.82.
\label{maxeta}
\end{equation}
\end{appendix}

\end{document}